\documentclass[notitlepage,letterpaper,12pt]{article}
\usepackage[ansinew]{inputenc} 
\usepackage[colorlinks=true,urlcolor=blue,linkcolor=blue,citecolor=blue]{hyperref} 
\usepackage{amsmath}
\usepackage{amsfonts}
\usepackage{amssymb}
\usepackage{graphicx}
\usepackage{subfig}
\usepackage{cancel}
\usepackage{booktabs}
\usepackage{color}
\usepackage{bigints}
\usepackage[top=1cm, bottom=2cm,outer=2cm, inner=2cm]{geometry}
\usepackage{amsthm}
\usepackage{lineno}

\begin{document}
\title {Acceptability Conditions and \\  Relativistic Anisotropic Generalized Polytropes}
\author{
 \textbf{Daniel Su\'arez-Urango}\thanks{\texttt{danielfsu@hotmail.com}} \\
\textit{Escuela de F\'isica, Universidad Industrial de Santander,  }\\ 
\textit{Bucaramanga 680002, Colombia}; \\
\textbf{Justo Ospino}\thanks{\texttt{j.ospino@usal.es}} \\
\textit{Departamento de Matem\'atica Aplicada and} \\
\textit{Instituto Universitario de F\'isica Fundamental y Matem\'aticas,}  \\ 
\textit{Universidad de Salamanca, Salamanca, Spain;} \\
\textbf{H\'ector Hern\'andez}\thanks{\texttt{hector@ula.ve}}
and \textbf{Luis A. N\'{u}\~{n}ez}\thanks{\texttt{lnunez@uis.edu.co}} \\
\textit{Escuela de F\'isica, Universidad Industrial de Santander,}\\ 
\textit{Bucaramanga 680002, Colombia} and \\
\textit{Departamento de F\'{\i}sica,} \\
\textit{Universidad de Los Andes, M\'{e}rida 5101, Venezuela.} \\
}

\maketitle

\begin{abstract}
This paper explored the physical acceptability conditions for anisotropic matter configurations in General Relativity. The study considered a generalized polytropic equation of state $P=\kappa {\rho}^{\gamma}+\alpha \rho -\beta$ for a heuristic anisotropy. We integrated the corresponding Lane-Emden equation for several hundred models and found the parameter-space portion ensuring the physical acceptability of the configurations. Polytropes based on the total energy density is more viable than those with baryonic density, and small positive local anisotropies produce acceptable models. We also found that polytropic configurations where tangential pressures are greater than radial ones are also more acceptable.  Finally, convective disturbances do not generate cracking instabilities. Several models emerging from our simulations could represent candidates of astrophysical compact objects.
\end{abstract} 

\section{Introduction}
\label{Introduction}
The polytropic equation of state (EoS) is one of the most common assumptions for modelling self-gravitating matter distributions in Newtonian and relativistic astrophysical scenarios. From the dynamics and stability of galaxies~\cite{Wolansky1999} to the description of the compact object's inner structure~\cite{Chandrasekhar1967, Tooper1964, Kovetz1967}, passing through mechanisms involved in stellar evolution~\cite{Fowler1964, RappaportVerbunt1983, CookShapiroTeukolsky1994}, this assumption has a long and venerable tradition in Astrophysics.

On the other hand, local anisotropy --non-Pascalian fluid description with unequal radial and tangential stresses, i.e.  $P \neq P_\perp$-- is also becoming a familiar premise.  Since the pioneering works of J. H. Jeans~\cite{Jeans1922}, and G. Lema\^itre~\cite{Lemaitre1933} this assumption is gaining momentum in describing Newtonian and relativistic matter configurations (see~\cite{Ruderman1972, BowersLiang1974, CosenzaEtal1981, HerreraNunez1989, HerreraSantos1997, MartinezRojasCuesta2003, HerreraBarreto2004, HerreraEtal2014, Setiawan2019} and references therein). Concerning local anisotropy, it is particularly interesting a recent paper~\cite{Herrera2020}, which discuss the instability of the isotropic pressure distribution in self-gravitating matter distribution.

Several heuristic strategies introduce anisotropy in relativistic matter configurations (see a detailed description in references~\cite{Ivanov2017, Ivanov2018}). Here we shall mention some of them. Firstly, the initial approach of Bowers \& Liang~\cite{BowersLiang1974}, followed by other schemes like: proportional to gravitation~\cite{CosenzaEtal1981};  quasilocal~\cite{DonevaYazadjiev2012};  covariant~\cite{RaposoEtal2019}; Karmarkar embedding class I~\cite{Karmarkar1948,OspinoNunez2020}; gravitational decoupling~\cite{Ovalle2017,AbellanEtal2020C}; double polytrope~\cite{AbellanEtal2020B}; conditioning the complexity factor~\cite{Herrera2018}  and finally, one of the most popular proposals: providing both, a particular barotropic equation of state $P= P(\rho)$ and a density profile (or equivalently a metric function)~\cite{Stewart1982, FinchSkea1989, HernandezNunez2004,  HerreraOspinoDiPrisco2008, HernandezNunez2013, AbellanEtal2020, NasheehaThirukkaneshRagel2021, HernandezSuarezurangoNunez2021}. Various of these strategies may lead to viable astrophysical models~\cite{Setiawan2019, BiswasBose2019, RahmansyahSulaksono2021}.

In a recent work~\cite{HernandezSuarezurangoNunez2021}, we considered the latter of the above approaches, i.e. introducing local anisotropy providing a polytropic EoS and an ansatz on the energy density profile. We found that this type of anisotropic matter distribution has a singular tangential sound velocity at the surface when the polytropic index is $n > 1$, and is commonly overlooked in the literature (see, for example, references~\cite{AbellanEtal2020, ThirukkaneshRagel2012, Ngubelanga2015,  TakisaMaharaj2013, Malaver2015, NgubelangaMaharaj2017, SharifSadiq2018, AzamNazir2020}). This is a general outcome when employing the ``standard'' polytropic EoS, $P~=~\kappa{\rho}^{1 + \frac{1}{n}}$, together with an ansatz on the metric functions.  It is worth mentioning that this pathology is not present in polytropes when any of the other strategies are implemented~\cite{CosenzaEtal1981, DonevaYazadjiev2012, RaposoEtal2019, Karmarkar1948, OspinoNunez2020, AbellanEtal2020C, AbellanEtal2020B}.

The recent detection of gravitational waves (GW)~\cite{AbbotEtalLigoCol2017, AbbottEtalLIGOVIRGCol2019}, and new results from X-ray astronomy~\cite{MillerEtal2019, MillerEtal2021} constrain the equation of state describing the Neutron Stars, NS, interiors. It has been crucial to identify, in the gravitational wave signal, the tidal deformation of the orbiting stars~\cite{FlanaganHinderer2008, Hinderer2008, BinningtonPoisson2009, DamourNagar2009} which reduces the parameter space for ultradense EoS~\cite{RadiceEtal2018, AbbottEtalLIGOVIRGCol2019}. Gravitational-wave astronomy provided an estimation of the tidal deformability of NS, while the Neutron star Interior Composition ExploreR, NICER, furnishes precise information about the mass and radius of selected pulsars~\cite{MillerEtal2021, MillerEtal2019}.

In this paper, we continue exploring models emerging from the generalised polytropic equation, $P=\kappa {\rho}^{\gamma}+\alpha \rho -\beta$, consisting of a combination of a polytrope plus a linear term~\cite{NasheehaThirukkaneshRagel2021, HernandezSuarezurangoNunez2021}. Based on this EoS, we follow the heuristic approach of L. Herrera and collaborators~\cite{CosenzaEtal1981} to include the anisotropic distribution of pressures within the matter configuration. We then integrate the corresponding Lane-Emden equation of stellar structure and identify the parameter space's portion, ensuring the acceptability conditions. We checked the stringent criteria of physical acceptability conditions put forth by B.V. Ivanov~\cite{Ivanov2017}, extended in reference~\cite{HernandezNunezVasquez2018}, and slightly improved in this work. 

We use this framework as a benchmark in looking for answers to the following questions:
\begin{itemize}
    \item Which are the most relevant parameters to ensure the acceptability for this type of anisotropic polytropes?
    \item Which anisotropy leads to more acceptable matter configurations: $\Delta_{+} = P_{\perp} - P > 0$ or $\Delta_{-} = P_{\perp} - P < 0$?
    \item Are low anisotropic models, $|\Delta | = | P_{\perp} - P| << 1$, more acceptable than those with high anisotropy, i.e. $|\Delta| = | P_{\perp} - P|  >> 1 $?
    \item Which of the relativistic polytropic EoS leads to a more acceptable distribution? Those based on the baryonic mass density or those implemented with the total energy density?
    \item Are these models consistent  with upper limit of the mass-weighted tidal deformability, $\tilde{\Lambda}_{(1.4)\star}$, predicted by LIGO~\cite{AbbottEtalLIGOVIRGCol2019} and the maximum mass limit discovered by NICER~\cite{MillerEtal2021}. 
\end{itemize}

This paper is organised as follows. Section \ref{FieldEquations} describes the notation and the framework of General Relativity. In Section \ref{PhysicalAcceptabilityConditions}, we list the set of acceptable conditions adhered to by our models to be acceptable compact stellar object candidates. In Section \ref{MasterLaneEmden}, we present the Lane-Emden anisotropic stellar structure equations to generalise a polytropic EoS. In Section \ref{NumericAcceptability} we discuss the modelling, explore the parameter space while fulfilling several of the acceptability conditions developed.  Finally, in Section \ref{FinalRemarks}, we present some final remarks and conclusions.

\section{The field equations}
\label{FieldEquations}
Let us consider the interior of a dense star described by  a spherically symmetric  space-time line element written as
\begin{equation}
\mathrm{d}s^2 = {\rm e}^{2\nu(r)}\,\mathrm{d}t^2-{\rm e}^{2\lambda(r)}\,\mathrm{d}r^2- r^2 \left(\mathrm{d}\theta^2+\sin^2(\theta)\mathrm{d}\phi^2\right),
\label{metricSpherical}
\end{equation}
with regularity conditions at 
$r=r_c=0$, i.e. ${\rm e}^{2\nu_c}=$ constant,  ${\rm e}^{-2\lambda_c}= 1$, and $\nu^{\prime}_c=\lambda^{\prime}_c=0$. 

Additionally, the interior metric should continuously match the Schwarzschild exterior solution at the sphere's surface,  $r=r_b=R$. This implies that ${\rm e}^{2\nu_b}={\rm e}^{-2\lambda_b}=1-2\mathcal{C}_{\star} = 1 -2M/R$,  where $M$ is the total mass and $\mathcal{C}_{\star}=M/R$ the compactness of the configuration.  From now on, the subscripts $b$ and $c$ indicate, respectively, the variable's evaluation at the boundary and the centre of the matter distribution.  

We shall consider a distribution of matter consisting of a non-Pascalian fluid represented by an energy-momentum tensor:
\begin{equation}
T_\mu^\nu = \mbox{diag}\left[\rho(r),-P(r),-P_\perp(r),-P_\perp(r)  \right] \,,
\label{tmunu}
\end{equation}
where $\rho(r)$ is energy density, with $P(r)$ and $P_\perp(r)$ the radial and  tangential pressures respectively. 

From the Einstein's field equations we obtain these physical variables in terms of the metric functions as
\begin{eqnarray}
\rho(r)&=& \frac{ {\rm e}^{-2\lambda}\left(2 r \lambda^{\prime}-1\right)+1 }{8\pi r^{2}}\,,\label{FErho} \\
P(r) &=&  \frac {{{\rm e}^{-2\,\lambda}}\left(2r \,\nu^{\prime} +1\right) -1}{8 \pi\,{r}^{2}}\,\label{FEPrad} \qquad \textrm{and} \\
P_\perp(r) &=&-\frac{{\rm e}^{-2\lambda}}{8\pi}\left[ \frac{\lambda^{\prime}-\nu^{\prime}}r-\nu^{\prime \prime }+\nu^{\prime}\lambda^{\prime}-\left(\nu^{\prime}\right)^2\right] \label{FEPtan} \,, 
\end{eqnarray}
where primes denote differentiation with respect to $r$. 

Now,  assuming  the metric function $\lambda(r)$ is expressed in terms of the Misner ``mass''~\cite{MisnerSharp1964} as 
\begin{equation}
 m(r)=\frac{r^2}{2}R^{3}_{232} \; \Leftrightarrow \; m(r)=4\pi \int ^r_0 T^0_0r^2\mathrm{d}r \; \Rightarrow e^{-2\lambda}= 1-\frac{2 m(r)}{r}\, ,
\label{MassDef} 
 \end{equation}
the Tolman-Oppenheimer-Volkoff equation -- i.e. $T^{\mu}_{r \; ; \mu}~=~0$, the hydrostatic equilibrium equation-- for this anisotropic fluid can be written as   
\begin{equation}
\frac{\mathrm{d} P}{\mathrm{d} r} = -\underbrace{(\rho +P)\frac{m + 4 \pi r^{3}P}{r(r-2m)}}_{F_{g}} + \underbrace{\frac{2}{r}\left(P_\perp -P \right)}_{F_{a}} \, ,
 \label{TOVStructure1}
\end{equation}
which together with
\begin{equation}
\label{MassStructure2}
\frac{\mathrm{d} m}{\mathrm{d} r}=4\pi r^2 \rho \,,
\end{equation}
constitute the relativistic stellar structure equations. 

Clearly, it is equivalent to solve the Einstein system (\ref{FErho})-(\ref{FEPtan}) or to integrate the structure equations (\ref{TOVStructure1})-(\ref{MassStructure2}). In the first case we obtain the physical variables $\rho(r)$, $P(r)$ and $P_\perp(r)$ given the metric functions $\lambda(r)$ and $\nu(r)$, while in the second approach  we integrate the structure equations (\ref{TOVStructure1})-(\ref{MassStructure2}) obtaining two barotropic equations of state, $P=P(\rho)$ and $P_{\perp}=P_{\perp}(P(\rho),\rho) \equiv P_{\perp}(\rho)$. 

These two EoS involving the radial and tangential pressures, together with the matching conditions --i.e. initial conditions for the system of first-order differential equations--, $P(R)=P_{b}=0$ and $m(R)=m_{b}=M$, lead to a system of differential equations for $\rho(r)$ which can be solved to obtain the inner structure of a self-gravitating relativistic compact object. 

As is well known, NS have been modelled for decades as Pascalian fluids, with isotropic pressure distribution. However, a considerable number of studies have shown that the pressures within compact objects could be anisotropic, i.e. non-Pascalian fluids with unequal radial and tangential pressures,  $\Delta\equiv P_\perp-P\neq 0$~\cite{BowersLiang1974, HerreraSantos1997, Setiawan2019, Herrera2020, RaposoEtal2019}. It can influence the stability of the compact object --inducing cracking or overturning--, its mass-radius ratio, or/and its maximum mass (see~\cite{Herrera1992, DiPriscoHerreraVarela1997, AbreuHernandezNunez2007b, GonzalezNavarroNunez2015, GonzalezNavarroNunez2017} and the corresponding bibliographies therein, particularly, references~\cite{HerreraSantos1997} and~\cite{Herrera2020}).  

As pointed out in the seminal work of Anisotropic spheres in General Relativity, by Bowers \& Liang~\cite{BowersLiang1974},  we distinguish two opposite terms in equation (\ref{TOVStructure1}) --the gravitational force, $F_{g}$, and the anisotropic strength, $F_{a}$-- which compete to shape the reacting pressure gradient. Clearly, the pressure steepness loosen when the anisotropy is positive, $\Delta_{+} = P_{\perp} - P > 0$, and tighten up if $\Delta_{-} = P_{\perp} - P < 0$. Thus, for a fixed central stiffness $\sigma = P_c/\rho_c$, the compactness, $\mathcal{C}_{\star}$, of the sphere increases for positive anisotropy $\Delta_{+}$ and decreases for negative anisotropy $\Delta_{-}$. In the first case, we can pack more massive configurations than the isotropic, $\Delta_{0} = 0$, occurrence, because the tangential stresses support the mass shells dropping the needed reacting radial pressure in these circumstances~\cite {CosenzaEtal1981, HerreraNunez1989, HerreraSantos1997}. If both forces balance, i.e. $F_{g}=F_{a}$, we obtain the particular matter configuration having vanishing radial pressures but only supported by tangential stresses~\cite{Florides1974}.

\section{The physical acceptability conditions}
\label{PhysicalAcceptabilityConditions}
In addition to solving the structure equations (\ref{TOVStructure1}) and (\ref{MassStructure2}) for a particular set of equations of state (e.g. $P=P(\rho)$ and $P_{\perp}=P_{\perp}(\rho)$), the emerging physical variables have to comply with the several acceptability conditions~\cite{DelgatyLake1998}. B.V. Ivanov~\cite{Ivanov2017, Ivanov2018} discussed the several independent acceptability conditions fulfilled by any relativistic anisotropic compact object. 

Acceptability conditions are crucial when considering self-gravitating stellar models. Only acceptable objects are of astrophysical interest and, in this work, those models have to comply with nine requirements stated as:
\begin{enumerate}
\item[{\bf C1:}] $2m/r < 1$, which implies
    \begin{enumerate}
    \item That the metric potentials $\textrm{e}^{\lambda}$ and $\textrm{e}^{\nu}$ are positive, finite and free from singularities within the matter distribution, satisfying $\textrm{e}^{\lambda_{c}} = 1$ and $\textrm{e}^{\nu_{c}}= \mbox{const}$ at the center of the configuration.
    \item The inner metric functions match the exterior Schwarzschild solution at the boundary surface.
    \item The interior redshift should decrease with the increase of $r$~\cite{Buchdahl1959,Ivanov2002B}. 
    \end{enumerate}

\item[{\bf C2:}] Positive density and pressures, finite at the center of the configuration with $P_c=P_{\perp c}$~\cite{Ivanov2002B}.

\item[{\bf C3:}] $\rho^{\prime} < 0$, $P^{\prime} < 0$, $P_{\perp}^{\prime} < 0$ with density and pressures having maximums at the center, thus $\rho^{\prime}_{c}=P^{\prime}_{c} = P^{\prime}_{\perp c}=0$ with $P_{\perp} \geq P$. 

\item[{\bf C4:}] The trace energy condition  $\rho - P - 2P_{\perp} \geq 0$, which is more retrictive than the strong energy condition, $\rho + P + 2P_{\perp} \geq 0$, for imperfect fluids~\cite{ Ivanov2018, KolassisSantosTsoubelis1998,PimentelLoraGonzalez2017}.

\item[{\bf C5:}]  The dynamic perturbation analysis restricts the adiabatic index~\cite{HerreraSantos1997,HeintzmannHillebrandt1975,ChanHerreraSantos1993,ChanHerreraSantos1994} 
\[
\Gamma = \frac{\rho + P}{P} v_s^{2} \geq \frac{4}{3} \,.
\]

\item[{\bf C6:}] The causality conditions on sound speeds: $0 < v_{s}^2 \leq 1$ and $0 < v_{s \perp}^2 \leq 1$~\cite{ AbreuHernandezNunez2007b, DelgatyLake1998}.

\item[{\bf C7:}] The Harrison-Zeldovich-Novikov stability condition: $\mathrm{d}M(\rho_c)/\mathrm{d}\rho_c > 0$~\cite{HarrisonThorneWakano1965,ZeldovichNovikov1971}.

\item[{\bf C8:}] The cracking instability against local density perturbations, $\delta \rho = \delta \rho(r)$, briefly described in Appendix \ref{appendix:a} (for more details the reader is referred to~\cite{HernandezNunezVasquez2018, GonzalezNavarroNunez2015, GonzalezNavarroNunez2017}).  

\item[{\bf C9:}] The adiabatic convective stability condition  $\rho^{\prime \prime} \leq 0$, which is more restrictive than the outward decreasing density and pressure profiles~\cite{HernandezNunezVasquez2018}.
\end{enumerate}

Notice that condition \textbf{C1} differs from condition $(m/r)^{\prime} > 0$ in~\cite{Ivanov2017}. The reasons for this change will be justified with a counterexample in Section \ref{ExaminingConditions}.

Observe that in references \cite{Ivanov2017} and \cite{Ivanov2018}, B.V. Ivanov  assumes in {\bf C3} that $P_{\perp} \geq P$ avoiding a global cracking perturbation instability~\cite{AbreuHernandezNunez2007b}. In principle, this assumption is not mandatory for anisotropic fluids but is commonly adopted in the literature because it allows more massive matter configurations. The requirement, $P_{\perp} \geq P$,  implies that the sign of the anisotropic force $F_a$ may counterbalance the gravitational force $F_g$. As we shall show, the anisotropic heuristic scheme chosen in the present work~\cite{CosenzaEtal1981} satisfies this condition for all models.

As discussed in references~\cite{DevGleiser2003} and \cite{GleiserDev2004}, the condition {\bf C5}, borrowed from the isotropic case, does not consider the complex behaviour of non-Pascalian fluids. There, M. Gleiser and K. Dev extended the formalism developed by Chandrasekhar to study the stability of general relativistic isotropic spheres against radial perturbations. They obtained stable relativistic anisotropic spheres having adiabatic exponents differing from the isotropic case.

In this work, our models incorporate in {\bf C8} a more elaborate cracking criterion considering local density perturbations, $\delta \rho~=~\delta \rho(r)$~\cite{HernandezNunezVasquez2018, GonzalezNavarroNunez2015, GonzalezNavarroNunez2017}. Local perturbed schemes are based on the fluid variables' reaction to a density fluctuation that drives the system out of its equilibrium. In this case,  pressure gradients may be affected by stabilizing the system.  (see Appendix \ref{appendix:a} and references~\cite{HernandezSuarezurangoNunez2021, HernandezNunezVasquez2018} for detailed discussions). 

Finally, the instability due to convection, {\bf C9}, has almost been forgotten in most stability analyses. It is an elementary criterion that implements the Archimedes principle in any hydrostatic matter configuration~\cite{Kovetz1967, Bondi1964B, Thorne1966}. 

\section{The Lane-Emden anisotropic equation of structure} 
\label{MasterLaneEmden}
The Lane-Emden equation is a dimensionless form of Tolman-Oppenheimer-Volkoff expression (\ref{TOVStructure1}) for a polytropic EoS~\cite{Tooper1964}. In this section, we shall derive the corresponding relativistic hydrostatic equilibrium equation for a generalized polytropic EoS~\cite{HernandezSuarezurangoNunez2021, NasheehaThirukkaneshRagel2021} (see Appendix \ref{appendix:b}). Following a heuristic strategy used in reference~\cite{HerreraBarreto2013}, we integrate it for a wide range of its parameter space, and the modelling performed will be discussed in Section \ref{NumericAcceptability}.

\subsection{The ``master'' Lane-Emden equation}
\label{MasterLaneEmdenParameter}
Just for completeness and to identify the physical parameters involved, we outline here the main characteristics of the master polytropic EoS,
\begin{equation}
    P(\rho) =\kappa {\rho}^{1 + \frac{1}{n}}+\alpha \rho -\beta  \,.
\label{MasterPolytropic}
\end{equation}
where $P$, $\rho$, $\kappa$ and $n$ are: the isotropic pressure, the mass  density and the  polytropic index, respectively (for details, we refer the reader to previous paper concerning this EoS~\cite{HernandezSuarezurangoNunez2021, NasheehaThirukkaneshRagel2021}). 

Notice that $\kappa, \alpha$ and $\beta$ are non-independent parameters. From equation (\ref{MasterPolytropic}), and the fact that on the surface the radial pressure vanishes, we have
\begin{equation}
  \beta=\kappa {\rho_b}^{1 + \frac{1}{n}}+\alpha \rho_b  \,,
  \label{betamasters}
\end{equation}
 with
\begin{equation}
\kappa=  
\frac{\sigma - \alpha\left[1 - \varkappa \right]}{ {\rho_c}^{\frac{1}{n}}\left[1 - \varkappa^{1 + \frac{1}{n}}\right] }
\,,
\label{kmaster}
\end{equation}
where
\begin{itemize}
    \item $\sigma = P_c/\rho_c$,  describes the stiffness at the centre of the matter distribution and
    \item $\varkappa =  \rho_b/\rho_c$, sketches the density drop from the centre to the surface of the compact object.
\end{itemize}

\begin{figure}[t!]
\centering
\includegraphics[width=2.9in,height=2.5in]{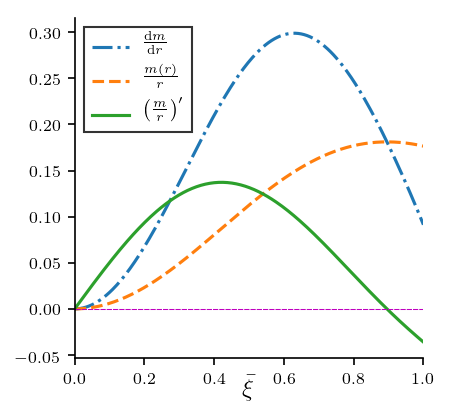}\hspace{1.0cm}
\includegraphics[width=2.9in,height=2.5in]{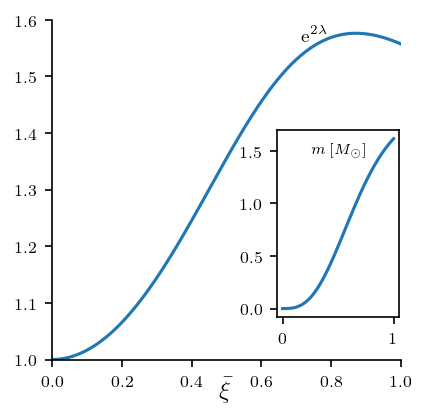}
\caption{The left plot displays $m^{\prime} \equiv \Upsilon (n+1)  \dot{\eta}$; $m/r \equiv \Upsilon (n+1) \eta / \xi$ and $(m/r)^{\prime} \equiv \Upsilon (n+1) (\dot{\eta}/\xi - \eta/\xi^{2})$, for models with $n=1.5$, $\sigma = 0.12$, $\alpha = -0.01$, $\varkappa = 0.05$ and $C =0.0625$. The right plot exhibits the metric potential $\mathrm{e}^{2\lambda} \equiv 1 - 2\Upsilon (n+1) \eta / \xi$ and the mass function in an inset plot. In this particular case, near the boundary of the configuration, we have $(m/r)^{\prime} < 0$ and a physically reasonable metric coefficient.  Thus, $(m/r)^{\prime} > 0$ should be considered as a sufficient but not a necessary condition.}
\label{mrderivative}
\end{figure}

To apply the master polytropic equation of state (\ref{MasterPolytropic}) in more realistic astrophysical scenarios, we integrated numerically the system of structure equations (\ref{TOVStructure1}) and  (\ref{MassStructure2}) assuming the equation of state (\ref{MasterPolytropic}). This lead to a generalization of the Lane-Emden equation for anisotropic relativistic fluids with the radial coordinate, energy density and mass written as
\begin{equation}
r=a\xi \,, \qquad  \rho=\rho_c \Psi^n(\xi) \, \quad \textrm{and} \quad  m= 4\pi a^3 \rho_c \ \eta(\xi) \, ,
\label{cambiovariable}
\end{equation}
respectively. Here $a$, is the ``Lane-Emden dimension radius'', which can be written in terms of the most fundamental physical parameters as
\begin{equation*}
    a^{2} = \frac{\Upsilon \left(1+n\right)}{4 \pi \rho_{c}} \quad \textrm{with} \quad  \Upsilon = \kappa \rho_c^{1/n} = \frac{\sigma - \alpha \left(1 - \varkappa \right)}{1 - \varkappa^{1 + 1/n}}\,.
\end{equation*}

Replacing the new variables in equation  (\ref{MasterPolytropic}) and dividing by central density,  we have
\begin{equation}
\mathcal{P} \equiv \frac{P}{\rho_c}= \Upsilon \left( \Psi^{n+1} - \varkappa^{1 + 1/n} \right) + \alpha \left(\Psi^{n} - \varkappa \right) \, .
\label{Padi}
\end{equation}
Now, with equation (\ref{Padi}) and considering the expressions (\ref{cambiovariable}),
 TOV equation (\ref{TOVStructure1}) can be expressed as
\begin{equation}
    \dot{\Psi}(\xi) = -\frac{1}{\xi}\left[  \frac{\left[\eta + \xi^{3} \mathcal{P} \right] \left[1 + \mathcal{P} \Psi^{-n} \right]}{ \xi - 2 \Upsilon \left(1+n\right) \eta  } - \frac{2 \Delta}{\rho_{c} \Upsilon \left(1+n \right) \Psi^{n}} \right] \left[1 + \frac{\alpha n}{\Upsilon (1+n) \Psi} \right]^{-1} \,,
    \label{TOVAdi}
\end{equation}
where, as usual, $\Delta= P_\perp-P$ represents the anisotropy. 

The second structure equation (\ref{MassStructure2}), becomes
\begin{equation}
    \dot{\eta} = \xi^{2}\Psi^{n} \,.
     \label{MassAdi}
\end{equation}
In both equations (\ref{TOVAdi}) and (\ref{MassAdi}), the dots denote derivatives with respect to the new variable $\xi$. Finally, we shall define $\bar{\xi}~=~\xi/\xi_b$ as a plotting device.

\begin{figure}[t!]
\centering
\includegraphics[height=5.0in,width=5.0in]{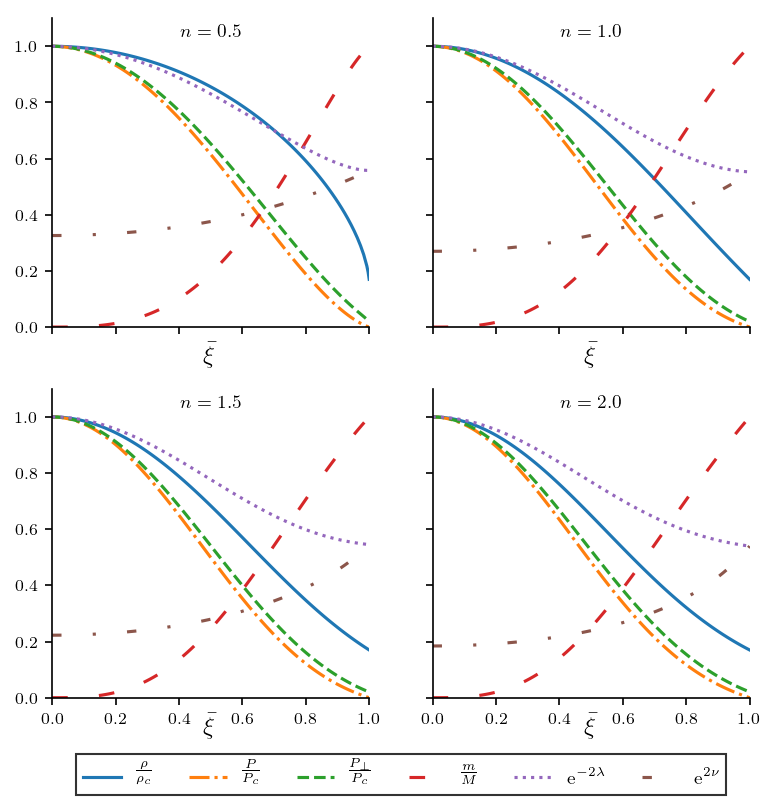}
\caption{Physical variables (density, radial/tangential pressures, mass) and metric coefficients within matter configuration for the numeric solution of the Lane-Emden master equation. They are displayed as functions of $\bar{\xi}$ for models having $\alpha = -0.01$, $\varkappa = 0.17$, $C = 0.05$, $\sigma = 0.175$, for several values of  $n$. The metric potentials are not singular. Density, radial and tangential pressures are decreasing functions of the $\bar{\xi}$ variable. All the physical and geometrical variables  for numeric models are well behaved and comply with the acceptability criteria {\bf C1}, {\bf C2} and {\bf C3}.}
\label{InternalStrucNumModels}
\end{figure}

\subsection{The anisotropic scheme and the Lane-Emden equation}
\label{NumericSolutions}
Several years ago, L. Herrera and W. Barreto, using a heuristic strategy, developed a general formalism to incorporate anisotropy in polytropic Newtonian and relativistic spheres~\cite{CosenzaEtal1981, HerreraBarreto2013, HerreraBarreto2013B}. Their heuristic procedure implements a previous method found in~\cite{CosenzaEtal1981}, which assumes the anisotropy proportional to the gravitational force, $F_g$, as
\begin{equation}
   \Delta\equiv P_{\perp} - P= C r F_g \equiv C r (\rho + P)\left[ \frac{m + 4 \pi r^3 P}{r(r-2m)}\right]\,, 
   \label{DelCosenza}
\end{equation}
where $C$ quantifies the anisotropy sign in each model. Replacing the former equation into  (\ref{TOVStructure1}) we get
\begin{equation}
    \frac{\mathrm{d}P}{\mathrm{d}r} = - h \frac{(\rho + P)(m + 4 \pi  r^3 P)}{r(r-2m)}\,,
\label{ansatzcosenza}
\end{equation}
with  $h = 1 - 2 C$. It is clear that when $h = 1$ the isotropic case is recovered. 

Notice that {\bf C3} condition and equation (\ref{ansatzcosenza}) implies $h > 0$, therefore if $\rho_{b} \neq 0$ we have, 
\begin{equation}
h = 1 - 2 C > 0 \,\, \Rightarrow \,\, C < \frac{1}{2} \,, \quad \text{and since }\quad  P_{\perp} \geq 0 \,\, \Rightarrow \,\, 0 \leq C < \frac{1}{2} \, .
\label{PtGreaterP}
\end{equation}
The tangential pressure should be positive at the boundary $P_{\perp \, b} \geq 0$ within the matter distribution, and from equation (\ref{DelCosenza}) it restricts the anisotropic parameter to  $0 \leq C < \frac{1}{2}$ for EoS having $\rho_{b} \neq 0$. This is the case for the generalised polytrope (\ref{MasterPolytropic}) with $\beta \neq 0$. When $\beta = 0$ (or any EoS that admits $\rho_{b} = 0$) the range of the anisotropy factor becomes $C < \frac{1}{2}$ admitting negative values for the $C$.

\begin{figure}[t!]
\centering
\includegraphics[height=2.8in,width=5.8in]{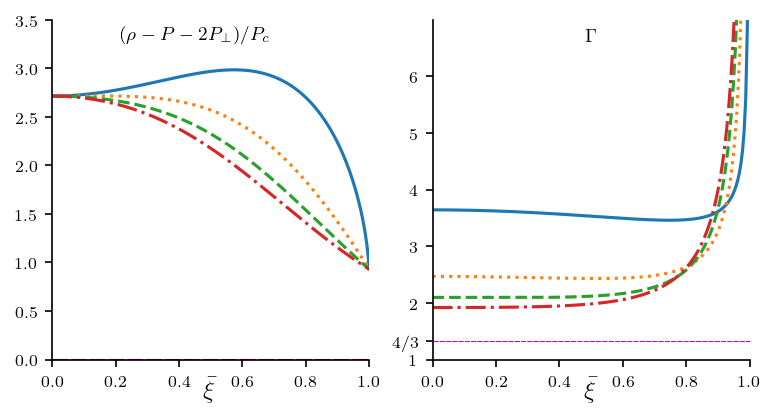} \\
\includegraphics[width=4.5in,height=0.3in]{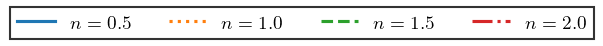}
\caption{Trace energy condition (left plate) and adiabatic index gamma (right plate) as a function of $\bar{\xi}$ for the Lane-Emden master equation, for models with parameters $\alpha = -0.01$, $\varkappa = 0.17$, $\sigma=0.175$ and $C = 0.05$ for several values of $n$. From these plots it is clear that the models considered comply both {\bf C4} and {\bf C5} criteria for various polytropic indexes.}
\label{fig:EnergyCondGammaVson}
\end{figure}

The equation  (\ref{ansatzcosenza}) in the new ``polytropic''  variables (\ref{cambiovariable}) becomes
\begin{eqnarray}
\dot{\Psi}(\xi)  = - \frac{(1-2C)  \left[ \eta + \xi^{3}\mathcal{P} \right] \left[1 +  \mathcal{P} \Psi^{-n} \right] }  {\xi \left[ \xi- {2 \Upsilon\,\left(1+n \right) \eta } \right] } \left[1+\frac{\alpha n}{\Upsilon(1+n) \Psi}\right]^{-1} \,,
\label{TOVAdih}
\end{eqnarray}
and  (\ref{DelCosenza}) can now be written as (see Appendix \ref{appendix:b} for details)
\begin{equation}
  {\tilde \Delta} \equiv \frac{\Delta}{\rho_c} =  \frac{C \Upsilon \left(1+n\right)\left(\eta + \xi^{3}\mathcal{P}\right)\left(\Psi^{n}+ \mathcal{P} \right)}{\xi - 2 \Upsilon \left(1+n \right) \eta} \,.
  \label{DelCosenzadi}
\end{equation}

If we introduce equations (\ref{TOVAdih}) and (\ref{DelCosenzadi}) in (\ref{TOVAdi})-(\ref{MassAdi}) we can integrate them, with the appropriate set of initial conditions:
\begin{equation}
\Psi_{c}\equiv \Psi (\xi = 0) = 1 \, , \quad \eta_c \equiv \eta (\xi = 0) = 0 
\end{equation}
and
\begin{equation}
   \mathcal{P}_{b}\equiv  \mathcal{P}(\xi = \xi_{b}) =  \Upsilon \left( \Psi^{n+1}_{b} - \varkappa^{1 + 1/n} \right) + \alpha \left(\Psi^{n}_{b} - \varkappa \right) = 0 \,. 
\end{equation}

\section{Modelling and acceptability conditions}
\label{NumericAcceptability}
In this section, we shall examine, through extensive modelling, the consequences of the acceptability conditions. We identify the most relevant EoS parameters, their range and their relevance in the specific acceptability conditions. We also explore the model stability associated with the sign of the anisotropic term, $\Delta = P_{\perp} - P$ in equation (\ref{TOVStructure1}).  Finally, we investigate the effect of the models' acceptability emerging from both relativistic polytropes considered.

\subsection{Examining the acceptability conditions for anisotropic models}
\label{ExaminingConditions}
The standard $2m/r < 1$ condition is different from the stronger $(m/r)^{\prime} > 0$, required by B.V. Ivanov in~\cite{Ivanov2017}. Clearly, if $(m/r)^{\prime} > 0$ we obtain well behaved metric functions but there are cases with $(m/r)^{\prime} < 0$ also having physically reasonable metric coefficients. One of these examples can be appreciated in Figure \ref{mrderivative} where, despite $(m/r)'~<~0$, the metric potential and physical variables fulfil all the required conditions {\bf C1a}, {\bf C1b} and {\bf C1c}. Thus, $(m/r)^{\prime} > 0$ should be considered as a sufficient but not a necessary condition. 
\begin{figure}[t!]
\centering
\includegraphics[height=2.8in,width=5.0in]{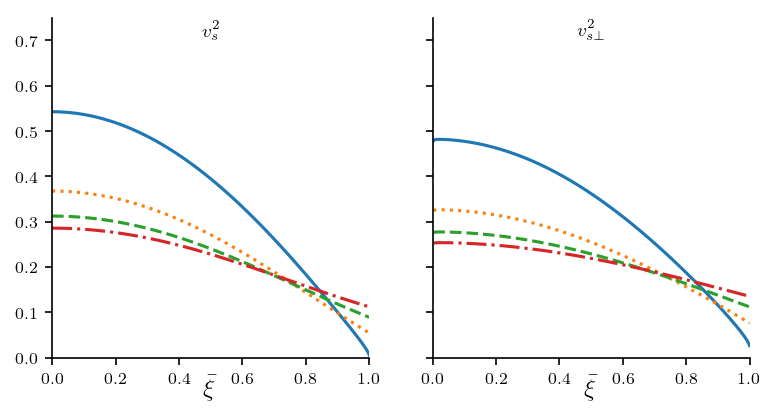} \\
\hspace{0.5cm}\includegraphics[width=4.5in,height=0.3in]{Figures/Labeln.png} 
\includegraphics[height=2.8in,width=5.0in]{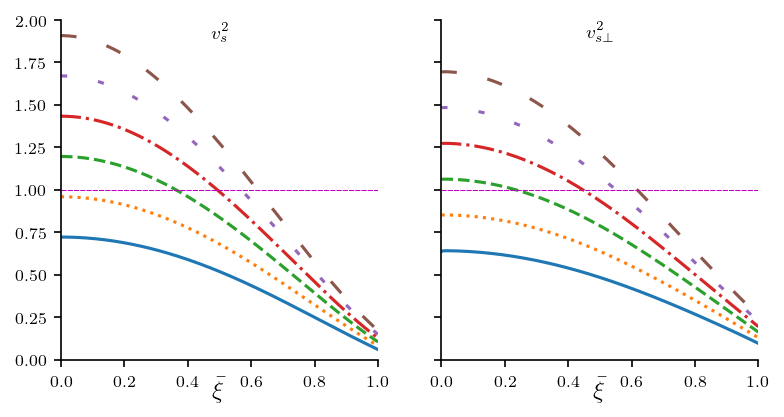}\\
\hspace{0.75cm}\includegraphics[width=3.5in,height=0.5in]{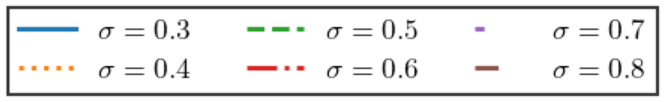} 
\caption{Radial sound speed (left plates) and tangential sound speed (right plates) as a function of $\bar{\xi}$ for the Lane-Emden master equation, for models with parameters $\alpha = -0.01$, $\varkappa = 0.17$ and $C = 0.05$. Top plates with $\sigma = 0.175$ and several values of $n$. Bottom plates with $n = 0.75$ and several values of $\sigma$. From these plots it is clear that the models considered comply causality condition  {\bf C6} for several polytropic indexes. High values of the stiffness $\sigma$ may have radial and tangential sound speed higher than the speed of light and are discarded.}
\label{fig:EnergyCondVson}
\end{figure}

On the other hand, Figure \ref{InternalStrucNumModels}  displays the metric coefficient and the physical variables profiles for various polytropic indexes. The metric potentials are not singular, and the physical variables are well behaved. Density, radial and tangential pressures are decreasing functions of the radial $\xi$ variable. Thus, we found the obvious restriction on the local compactness, $\mathcal{C}=m/r<1/2$ is enough to obtain physically reasonable metric coefficients. In Section \ref{SignStrengthAnisotropy} we shall discuss in details the implementation of {\bf C3} with the restriction (\ref{PtGreaterP}).

Figure \ref{fig:EnergyCondGammaVson} (left plate) displays the fulfilment of the trace energy condition {\bf C4}, as well as the expected dynamic stability criterion {\bf C5}, for models having different polytropic indices ($n=0.5$ through $n=2.0$) and stiffness $\sigma$. We will use criterion {\bf C4} to limit the possible values of $\sigma$, so that with $0<\sigma<1/3$ we guarantee the fulfilment of conditions {\bf C1} - {\bf C9}.
\begin{figure}[t!]
\centering
\includegraphics[height=5.0in,width=5.0in]{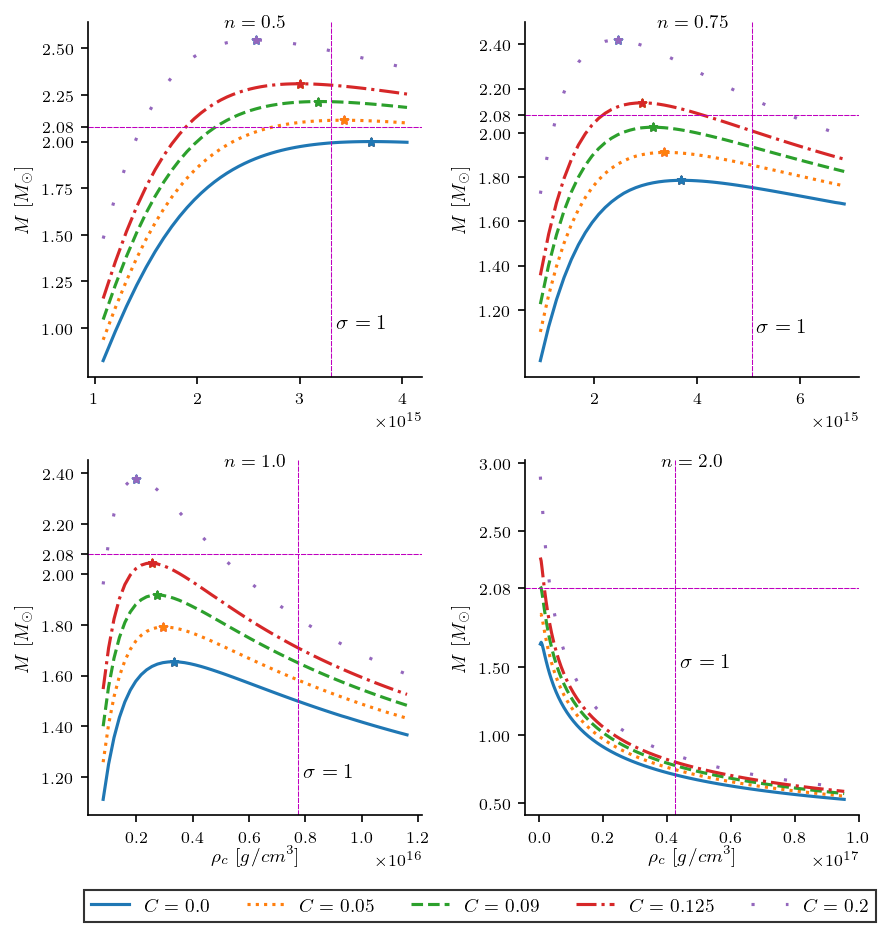}
\caption{The total mass, $M$, as a function of $\rho_c$ for models with parameters $\alpha = -0.01$, $\varkappa = 0.17$, and varying $C$ (the anisotropic factor) for several values of the polytropic index $n$. From the graphics, we can see that the shape of the curves are highly related to the polytropic index $n$, and they are susceptible to $C$. The smaller the polytropic index $n$ is, the greater the range of densities that fulfil {\bf C7}. A star-shaped mark represents maximum masses, which dictate the onset of unstable models. In each plot, have set two lines. One horizontal indicating the $2.08~M_{\odot}$ recently discovered pulsar~\cite{MillerEtal2021} and other vertical representing the stiffest condition at the centre of the distribution, $\sigma = 1$. Several stable anisotropic models could describe masses like reported for J0740+6620 pulsar. }
\label{HZNCriterionMass}
\end{figure}

As we have pointed out, {\bf C5} does not consider the complex behaviour of non-Pascalian fluids. Even the formalism presented in references~\cite{DevGleiser2003, GleiserDev2004} assumes radial perturbations do not affect the tangential pressure distributions.  Figure \ref{fig:EnergyCondGammaVson} exhibits a singularity of the adiabatic index at the boundary surface. This effect emerges from the isotropic definition of $\Gamma$ implemented for an anisotropic EoS (\ref{MasterPolytropic}). Non-Pascalian fluids should have complex EoS among the state variables (energy density, radial/tangential pressures and other variables of state)~\cite{HorvatIIijicMarunovic2011B}. The validity of this criterion should be further explored when considering relativistic anisotropic matter distributions. 

It is also clear from Figure \ref{fig:EnergyCondVson}, that the models considered comply with the causality conditions on sound speeds, {\bf C6}; i.e.  $0 < v_{s}^2 \leq 1$ and $0 < v_{s \perp}^2 \leq 1$. As expected, the higher the stiffness, $\sigma$, is, the more restricted the models are. If we refer only to condition {\bf C6}, then those models having $\sigma \geq 2/3$ present a non-causal region near the centre of the matter distribution. On the other hand, models with lower $\sigma$, comply the criterion {\bf C6} for different values of the polytropic indexes. 
\begin{figure}[t!]
\centering
\includegraphics[height=5.0in,width=5.0in]{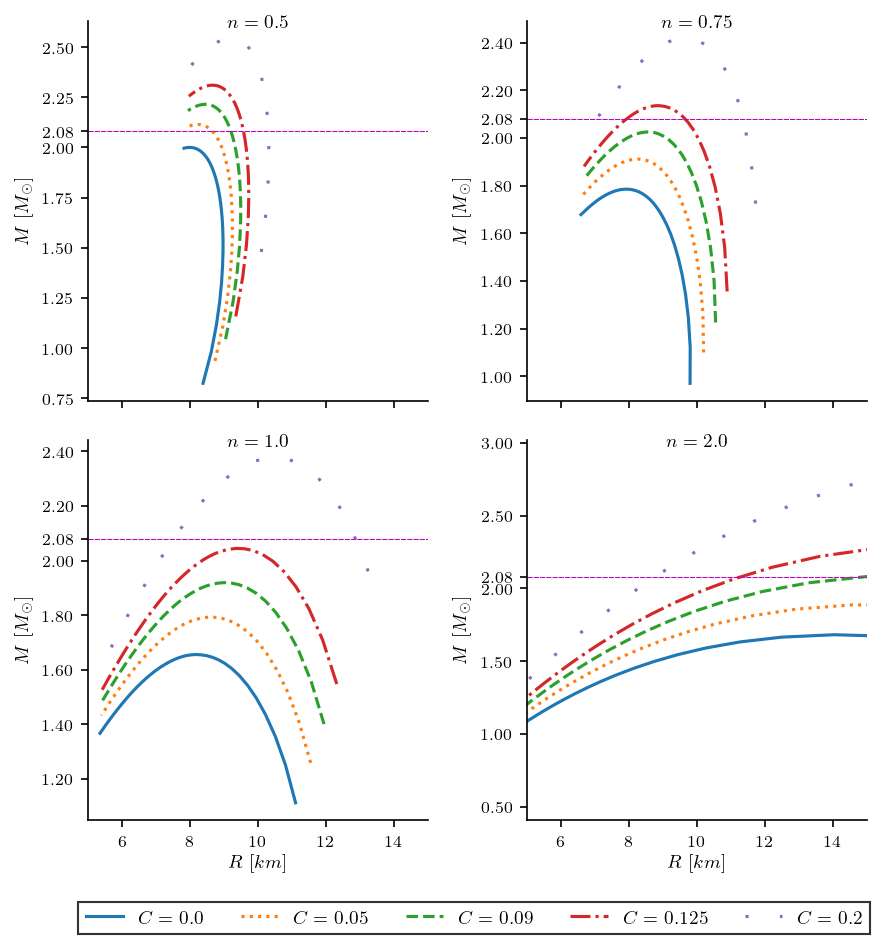}
\caption{Total mass-Total radius curves, $M-R$, for models with $\alpha = -0.01$, $\varkappa = 0.17$, and varying $C$ (the anisotropic factor) for several values of the polytropic index $n$. The curves appearance are highly related to the polytropic index $n$, and are sensitive to variations in the anisotropic index $C$. As in figure \ref{HZNCriterionMass} it is clear that, for the parameters $\alpha$, $\varkappa$ and $n$ considered here, only anisotropic models could describe $2.08~M_{\odot}$ NS~\cite{MillerEtal2021}.
}
\label{HZNMR}
\end{figure}

Configurations with positive anisotropy ($\Delta > 0$) allow more massive stellar models than isotropic ones (see section \ref{SignStrengthAnisotropy}). Thus, the Harrison-Zeldovich-Novikov stability condition {\bf C7}, ${\mathrm{d}M(\rho_c)/\mathrm{d}\rho_c > 0}$, is very sensitive to the anisotropic parameter $C$. For configurations with the same central density, an increase in anisotropy leads to an increment in the total acceptable mass, as seen in Figure \ref{HZNCriterionMass}. Therefore, anisotropic configurations need lower central densities to achieve stable models with the same total mass. $M(\rho_c)$ and $M-R$ curves (Figures \ref{HZNCriterionMass} and \ref{HZNMR} respectively) were set up varying $\sigma$ between 0.1 and 0.9 in steps of 0.025. When $n = 0.5$  lines close to the isotropic condition $ C = 0 $, do not climb to a maximum mass in this range. 

Criterion {\bf C8} concerns to the cracking instability for local density perturbations and Figure \ref{figCrackingPlot}, left panel, plots $\delta \mathcal{R}$ several polytropic indexes with fixed  $\alpha = -0.01$, $\varkappa = 0.17$, $\sigma=0.175$ and $C = 0.05$.  All these models satisfy the stability criterion {\bf C8} because no cracking or overturning occurs within the matter configuration. As discussed in references~\cite{ HernandezNunezVasquez2018, GonzalezNavarroNunez2015} and \cite{GonzalezNavarroNunez2017}, if the pressure gradient is not affected by the density perturbation, $\delta \mathcal{R}$ may change its sign, and potential cracking instabilities may appear. On the other hand, if the gradient reacts to the perturbation, we find that $\delta \mathcal{R}$ does not change sign, and the matter configuration becomes stable to cracking. The inset plot displays the effect of the global perturbation when the pressure gradient is not affected by the density perturbation. The change of sign for the force distribution near the boundary of the configuration is clear.

Finally, the convective adiabatic stability criterion, {\bf C9}, completes the set of acceptability conditions. As mentioned above, this is a simple criterion that implements the Archimedes principle introducing a very stringent condition in the density profile, i.e. $\rho^{\prime \prime} \leq 0$. Figure \ref{figCrackingPlot}, right panel, displays the convective adiabatic stability criterion, {\bf C9}. Only one of the models presented, $n~=~0.5$, fulfils this criterion within the whole configuration.

\subsection{The modelling, the acceptability and the parameter space}
\label{ParamenterSpace}
We shall identify the most significant parameters in the coming sections and relate them to the specific criterion's fulfilment (or failure).  We will also explore which anisotropic signatures lead to more acceptable models.  Finally, we explore the acceptability of the particular expressions for relativistic polytropes.

\subsubsection{The most significant acceptability parameters}
As can be seen from equations (\ref{MasterPolytropic}), (\ref{betamasters}), (\ref{kmaster}) and (\ref{DelCosenza}), there are five fundamental physical parameters for the generalized polytropic EoS (\ref{MasterPolytropic}), i.e.
\begin{itemize}
    \item $n$, the polytropic index
    \item $\alpha$, the linear coefficient related to the radial sound velocity 
    \item $\sigma = P_c/\rho_c$, the stiffness at the centre of the matter distribution,
    \item $C$, the anisotropic factor and     
    \item $\varkappa =  \rho_b/\rho_c$, the density drop from the centre to the surface of the compact object. 
\end{itemize}
\begin{figure}[t!]
\centering
\includegraphics[width=2.9in,height=2.9in]{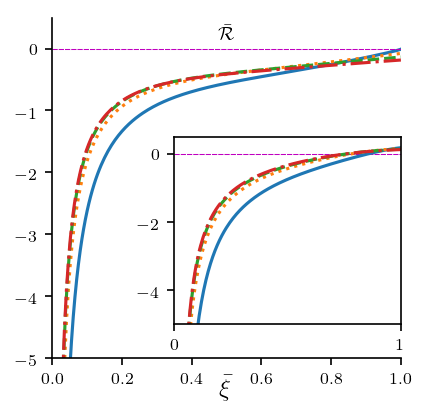}\hspace{1.0cm}
\includegraphics[width=2.9in,height=2.9in]{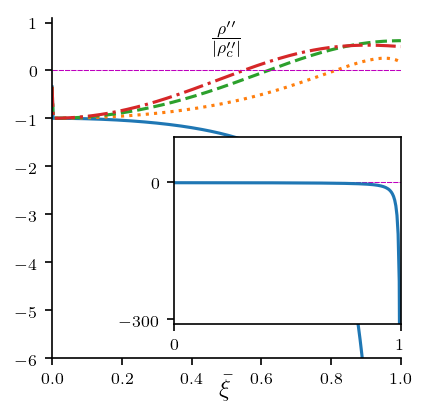} \\
\hspace{0.5cm}\includegraphics[width=4.5in,height=0.3in]{Figures/Labeln.png}
\caption{Perturbed hydrostatic equilibrium equation (left plate) and convective adiabatic stability criterion (right plate) as a function of $\bar{\xi}$ for the Lane-Emden master equation, for models with parameters $\alpha = -0.01$, $\varkappa = 0.17$, $\sigma=0.175$, $C = 0.05$ and several values of $n$. There are no changes in the sign of $\bar{\mathcal{R}}$ in the left plate. Therefore there is no cracking or overturning within the material configuration: all models shown fulfil {\bf C8}. The left plate's inset shows the effect when the local perturbations do not affect the pressure gradient. In this case, there is cracking near the boundary of the distribution. On the other hand, notice that only one of the models presented, $n~=~0.5$, fulfils the convective adiabatic stability criterion, {\bf C9}, within the whole configuration.}
\label{figCrackingPlot}
\end{figure}
We found that the first four parameters are the most significant because their variation gives us a wide range of acceptable models. 

The variation of $n$ describes a wide range of materials. For standard polytropic EoS ($\alpha= \beta =0$ in equation (\ref{MasterPolytropic})), the case $n = 0$ is associated with an incompressible fluid~\cite{Bludman1973}, while $n = 3$ is used to model an utterly degenerate gas in the relativistic limit~\cite{Horedt2004}.

The parameter $\sigma$ --the ratio of pressure to energy density at the centre of the configuration-- indicates the material's stiffness and how relevant the relativistic regime is. In case of $\sigma \rightarrow 0$ TOV equation (\ref{ansatzcosenza}) reduces to the Newtonian hydrostatic equilibrium equation~\cite{Tooper1964}, changing to the non-relativistic description of the fluid. 

\begin{figure}[t!]
\centering
\includegraphics[width=1.8in]{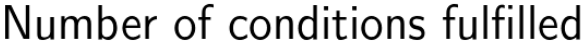} \\
\includegraphics[width=3.0in]{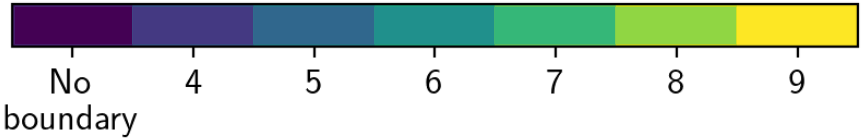} \\
\includegraphics[height=1.8in,width=5.0in]{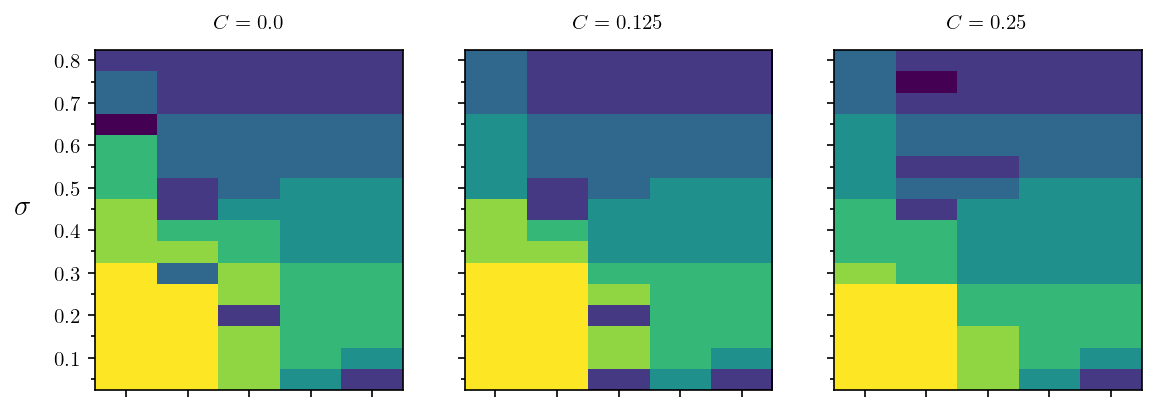}\\
\includegraphics[height=1.95in,width=5.0in]{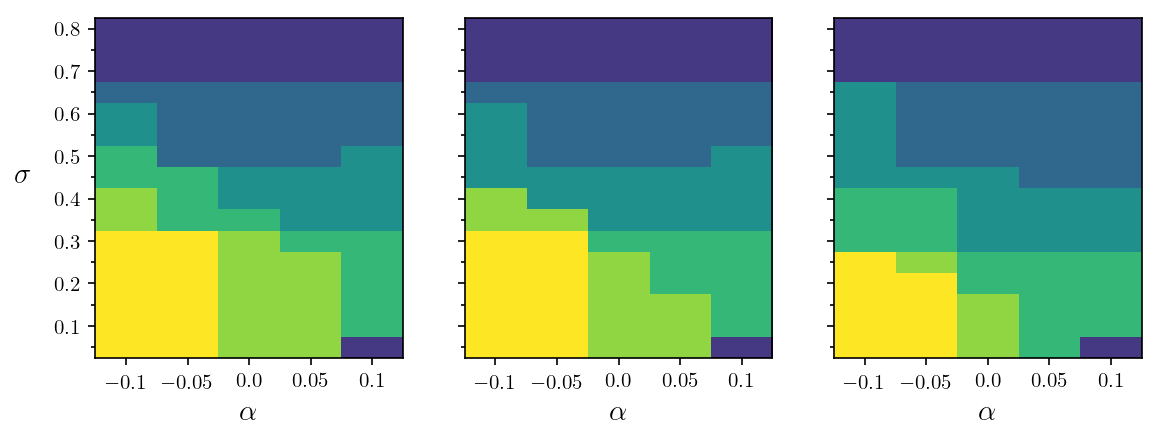}\\
\caption{Parametric space for the Lane-Emden master equation, for models with polytropic index $n = 1.0$ and parameters $\alpha$, $\sigma$ and $C$ varying from $-0.1$ to $0.1$, $0.05$ to $0.8$ and $0.0$ to $0.25$, respectively. When $\varkappa = 0.05$ (top plates) the parameter space does not differ much from the dataset implemented with $\varkappa = 0.2$ (bottom plates). Thus, the ratio between central and surface density does not have a greater incidence in parameter space variation.}
\label{ParSpac}
\end{figure}

As mentioned before, the linear coefficient $\alpha$ in master polytropic EoS is closely related to the speed of sound: positive values of $\alpha$ decrease the radial and tangential velocity of sound, while negative values have the opposite effect. However, models with negative values are more stable (see Figure \ref{ParSpac}).    

Finally, the parameter $\varkappa$ --the ratio between the central and the surface density-- does not have a greater incidence in parameter space variation. When $\varkappa = 0.05$ (Figure \ref{ParSpac}, top plates) the parameter space does not differ much from the data-set implemented with $\varkappa = 0.2$ (Figure \ref{ParSpac}, bottom plates). However, this parameter is important when dealing with convective stability.

The following section will show how the acceptability conditions are affected or constrained to particular ranges of these physical parameters.

\subsubsection{Parameters and the acceptability criteria}
\begin{figure}[t!]
\centering
\includegraphics[width=2.9in,height=2.9in]{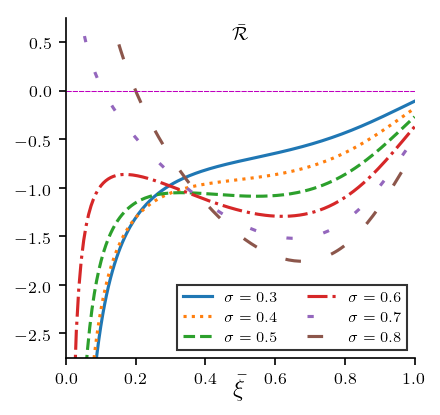} \hspace{1.0cm}
\includegraphics[width=2.9in,height=2.9in]{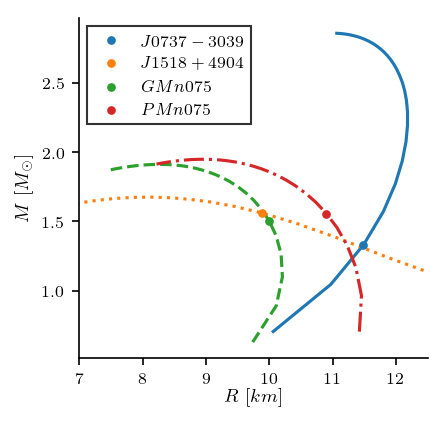}
\caption{Left: Perturbed hydrostatic equilibrium equation as a function of $\bar{\xi}$ for the Lane-Emden master equation, showing the distribution of the net force immediately after it departures from equilibrium, for models with parameters $\alpha = -0.01$, $\varkappa = 0.17$, $n=0.75$, $C = 0.05$ and several values of $\sigma$. Models that present cracking also unfulfill causality condition \textbf{C6} (see Figure \ref{fig:EnergyCondVson} bottom plates). 
Right: $M-R$ (Total mass-Total radius) curves for NS candidates in Table \ref{tableparameters1}. Central density increases from right to left along the curve. All candidates are in the stable region models (to the right of the maximum mass).}
\label{FigFracturaMRModels}
\end{figure}

For each criterion, we can associate an acceptability range for, at least, one of the above physical parameters:

\begin{itemize}
    \item \textbf{C4}: SEC fails when $\sigma$ is less than $1/3$. This comes straightforward from dividing SEC by $P_{c}$ and evaluating at the centre of the distribution.
    
    \item \textbf{C6}: Causality condition breaks down when the stiffness $\sigma$ is high. Figure \ref{fig:EnergyCondVson} (bottom plates)  shows how radial and tangential sound speeds increase proportionally to $\sigma$. Moreover, negative values of $\alpha$ also increase sound speed. This behaviour is characteristic for any value of the polytropic index $n$.
    \item \textbf{C7}: Polytropic index $n$ shapes the $M = M(\rho_c)$ curve  for Harrison-Zeldovich-Novikov criteria. The bigger $n$ is, the fewer models fulfil this condition. These plots are also sensitive to the variation of anisotropic factor $C$ in the same way as $n$ (shown in Figure \ref{HZNCriterionMass}). It is worth mentioning that in Figure \ref{HZNCriterionMass}, the central density values are given after the integration and do not affect the outline of the curves, only shifting them. 
    \item \textbf{C8}: Matter configurations may present cracking for higher $\sigma$ (see Figure \ref{FigFracturaMRModels}). However, these stiff models do not comply with the causality condition \textbf{C6}
    \item \textbf{C9}: Convective instabilities occur when the polytropic index $n$ is greater than $1$ (Figure \ref{figCrackingPlot}). However, a high value of $\varkappa$ may lead to the second derivative of $\rho$ being entirely concave since the density drop from the centre to the surface is low.
\end{itemize}

\subsubsection{The anisotropy and acceptability}
\label{SignStrengthAnisotropy}
It is clear that when $P_{\perp}>P$, a repulsive anisotropic force, $F_{a}$, appears in the hydrostatic equilibrium equation (\ref{TOVStructure1}) in the opposite direction to the gravitational force, $F_{g}$. Now, when $P_{\perp}<P$, implies that, both forces, $F_{a}$ and $F_{g}$, have the same direction~\cite{GokhrooMehra1994, DevGleiser2003}. In principle, both signs are permitted. However, usually, we find the preferred sign $P_{\perp}>P$ in the literature because it leads to more massive matter configurations, which could help to explain recent observations of high mass pulsars~\cite{MillerEtal2021, RileyEtal2019}. We show several of these examples in Table \ref{tableparameters1} at the end of the present section.

As we showed in equation (\ref{PtGreaterP}), the anisotropic heuristic scheme chosen in the present work~\cite{CosenzaEtal1981} satisfies the condition $P_{\perp} \geq P$ for all models with $\rho_{b} \neq 0$. Clearly, $P_{\perp} \geq P$ does not necessarily implies $P^{\prime}_{\perp} > P^{\prime}$, but the reverse is true. As stressed in \cite{Ivanov2017, Ivanov2018}, 
\begin{equation}
 P^{\prime}_{\perp} > P^{\prime} \,\, \Rightarrow \,\, v_{s_\perp}^2 < v_s^2 \,\, \Rightarrow \,\, P_{\perp} > P \, .
\label{PtPconsequences}
\end{equation}
In this case, $\delta \mathcal{R}_a$, the anisotropic force distribution emerging from the perturbation of the hydrostatic equation will always be in the same direction of the gravitational force $\delta \mathcal{R}_g$, (equations (\ref{Fp_pg}) and (\ref{PerturbAnisotropy}) in Appendix \ref{appendix:a}). If the density perturbations do not affect the pressure gradient, no cracking instability will appear. 
\begin{figure}[t!]
\centering
\includegraphics[width=1.8in]{Figures/LabelNumberFulfilled.png} \\
\includegraphics[width=3.0in]{Figures/LabelParSpace3.png} \\
\includegraphics[height=2.2in,width=5.6in]{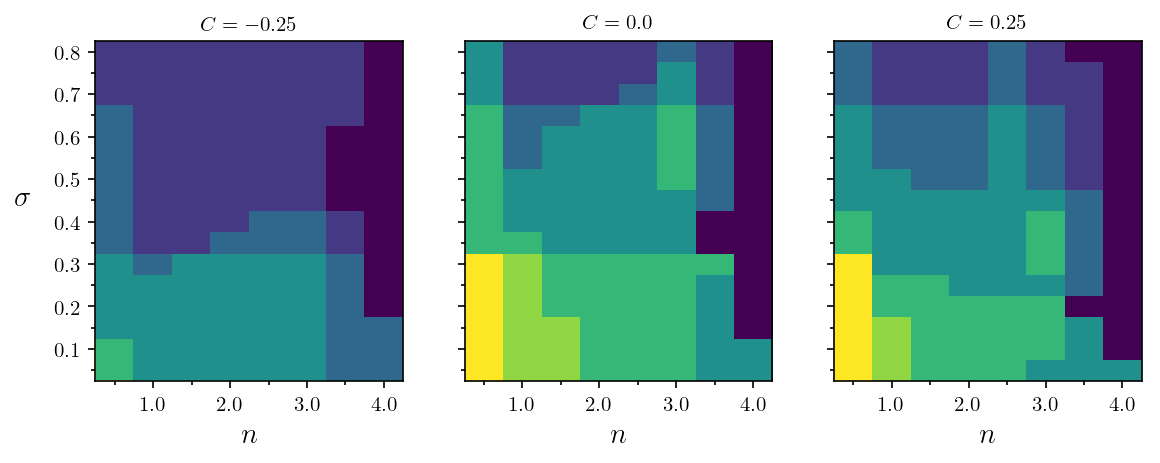}\\
\includegraphics[height=2.2in,width=4.0in]{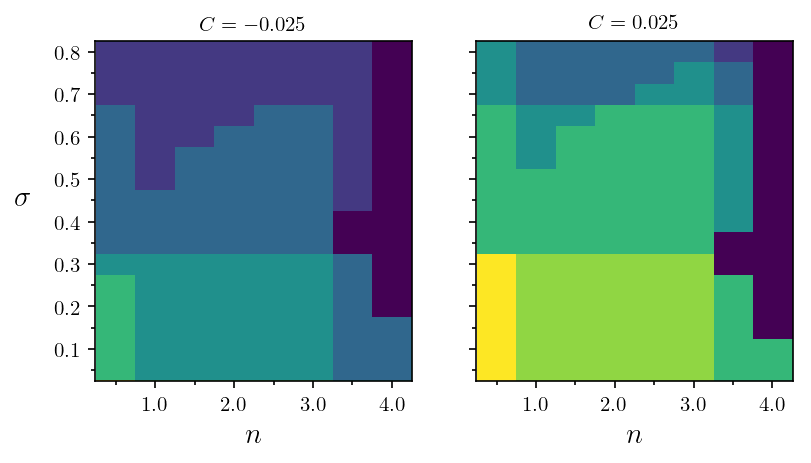}
\caption{Parametric space for the Lane-Emden equation $(\alpha = \beta = 0)$ with energy density (bottom plates in Figure \ref{figParSpaLE}), for several values of anisotropic factor $C$. Isotropic case, $C = 0.0$, corresponds to Tooper's     solutions~\cite{Tooper1964}, while $ C = 0.25$ corresponds to Herrera and Barreto's models \cite{HerreraBarreto2013}. In general, more stable models are obtained for small and positive values of the anisotropic factor.}
\label{Slices}
\end{figure}

On the other hand, for a local cracking perturbation approach, the value of the perturbation force should compete (or cooperate) with the magnitude of the reaction to the pressure gradient $\delta \mathcal{R}_p = \delta P^{\prime}$, creating cracking instabilities within the matter configuration. 

Finally, Figure \ref{Slices} explores in more detail the amount of anisotropy needed to generate stable models around the isotropic case.  We studied the perturbation of the isotropy for the standard polytropic EoS (i.e. $\alpha= \beta =0$ in equation (\ref{MasterPolytropic})). In this case we also reproduced two previous polytropes: isotropic Tooper's solutions for , $ C = 0.0$~\cite{Tooper1964}, and Herrera and Barreto's models for $ C = 0.25$~\cite{HerreraBarreto2013}. We considered both cases: $0~<C~<\frac{1}{2}~\Leftrightarrow~P_{\perp}>P$ and $C<0~\Leftrightarrow P_{\perp}<P$ and found that small positive anisotropy leads to more acceptable models.

\subsubsection{Energy density: baryonic {\it vs} total mass}
From the perspective of General Relativity, two formulations exist for polytropic EoS having the same Newtonian limit and only differing in the density considered: energy density or baryonic mass density. This difference could have significant consequences when describing the parameter space range in a compact object description~\cite{ArrollochavezEtal2020}. 

Following~\cite{HernandezSuarezurangoNunez2021, HerreraBarreto2013}, we briefly present both cases:
\begin{enumerate}
    \item First, we consider the particle density, $\mathcal{N} =\hat{\rho} / m_0$ with $m_0$ the baryonic mass and $\hat{\rho}$ the baryonic mass density and  combining the equation of state $P = \kappa \hat{\rho}^{1 + 1/n}$ with the adiabatic first law of thermodynamics, we obtain:
\begin{equation}
 \mathrm{d}\left(\frac{\rho}{\mathcal{N}} \right) + P\mathrm{d}\left(\frac{1}{\mathcal{N}} \right) = 0 \,\, \Rightarrow \,\,
   \frac{\mathrm{d}}{\mathrm{d}{\hat \rho}} \left(\frac{\rho}{\hat \rho}\right) =
\frac{P}{ {\hat \rho}^{2}} \quad \Rightarrow 
\frac{1}{{\hat \rho}}\frac{\mathrm{d} \rho}{\mathrm{d}{\hat \rho}}-\frac{\rho}{{\hat \rho}^2}= \frac{P}{ {\hat \rho}^{2}} \,,
\label{firstlaw1}
\end{equation}
thus, equation (\ref{firstlaw1}) can be integrated and we obtain two possible solutions 
\begin{equation}
\frac{\mathrm{d} \rho}{\mathrm{d}{\hat \rho}}-\frac{\rho}{{\hat \rho}} =    \kappa{\hat \rho}^{\gamma-1}   \,,
\quad \Rightarrow 
\left\{
\begin{array}{lllll}
\gamma \neq 1   &  \,\, \Rightarrow \,\, &\rho = \dfrac{\kappa {\hat \rho}^\gamma }{\gamma-1}+\varsigma_{1}{\hat \rho} & \\
                &    &  &   \\
\gamma = 1  &  \,\, \Rightarrow \,\,  & \rho = \left[\kappa \ln({\hat \rho}) +\varsigma_{1}\right]{\hat \rho}&
\end{array}
\right.
\label{firstlaw1b}
\end{equation}
where $\gamma=1 + \frac{1}{n}$ is the polytropic exponent, and $\varsigma_{1}$ a constant of integration.

\item The second approach takes into account the energy density $\rho$ and beginning with 
\begin{equation}
    P = \kappa \rho^{\gamma} \,,
\label{politroRho}
\end{equation}
so that equation (\ref{firstlaw1}) becomes
\begin{equation}
\frac{\mathrm{d} \rho}{\mathrm{d}{\hat \rho}}-\frac{\rho}{{\hat \rho}} = \frac{\kappa}{{\hat \rho}}\ {\rho}^{\gamma} \quad
\Rightarrow 
    \int \frac{\mathrm{d} \rho}{\kappa{ \rho}^{\gamma}+\rho}=\ln{\left(\frac{{\hat \rho}}{\varsigma_{2}}\right)} \qquad \textrm{with} \quad \gamma \neq 1 \,, 
\end{equation}
which can solved as  
\begin{equation}
    \rho= \frac {{\hat \rho}}{\varsigma_{2}} \left[ 1-\kappa\left(\frac {{\hat{\rho}}}{\varsigma_{2}}\right)^{\frac{1}{n}}
 \right]^{-n} = \frac{{\hat{\rho}}}{\left[ \varsigma_{2}^{\frac{1}{n}} 
 -\kappa {\hat{\rho}}^{\frac{1}{n}}\right]^n} \,,
\end{equation}
and if $\gamma= 1$, equation (\ref{firstlaw1}) can be integrated as
\begin{equation}
\frac{\mathrm{d} \rho}{\mathrm{d}{\hat \rho}}-\frac{\rho}{{\hat \rho}} = \frac{\kappa}{{\hat \rho}}\ {\rho} \quad
\Rightarrow 
\rho= \varsigma_{2} \, {\hat{\rho}}^{1 +\kappa} \,,
\end{equation}
again, with $\varsigma_{2}$ a constant of integration.
\end{enumerate}

\begin{figure}[t!]
\centering
\includegraphics[width=1.8in]{Figures/LabelNumberFulfilled.png} \\
\includegraphics[width=3.0in]{Figures/LabelParSpace3.png} \\
\includegraphics[height=1.8in,width=5.0in]{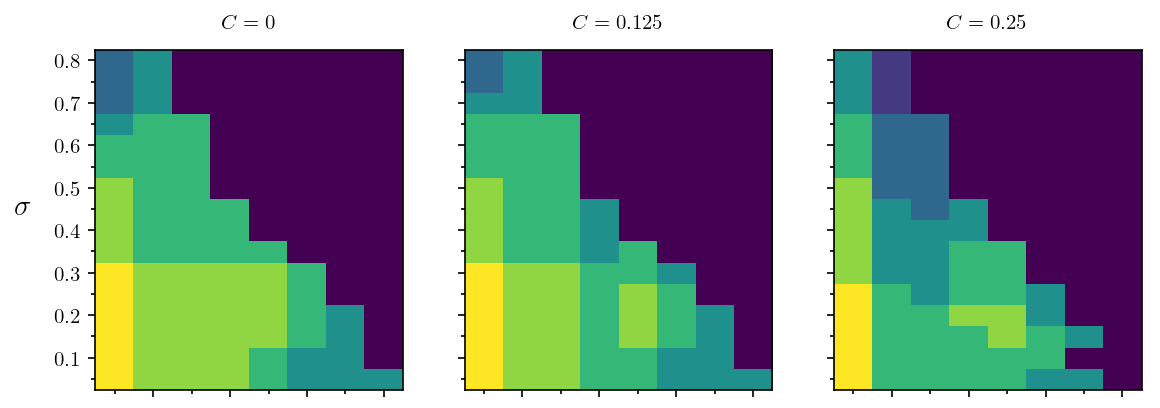}\\
\includegraphics[height=1.95in,width=5.0in]{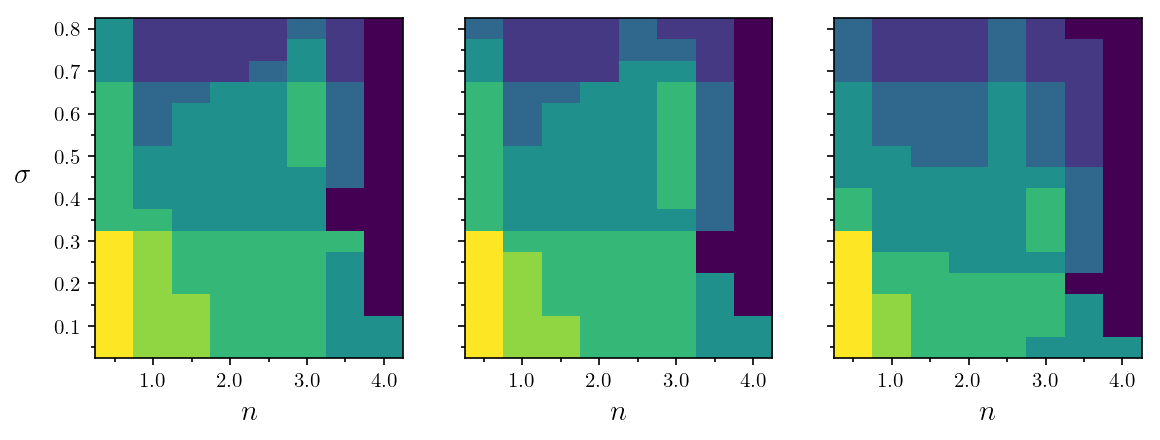}\\
\caption{Parametric space for the Lane-Emden equation $(\alpha = \beta = 0)$ with mass density (top plates) and energy density (bottom plates), for models with polytropic index from $n=0.5$ to $n=4.0$, and parameters $C$ and $\sigma$ varying from $0$ to $0.25$ and $0.05$ to $0.8$, respectively. None of the models with polytropic EoS with baryonic mass density fulfil all the conditions, and about half were not integrable.}
\label{figParSpaLE}
\end{figure}

As shown in Figure \ref{figParSpaLE}, the matter configurations are more viable for the total energy density, $\rho$ than those for baryonic mass density $\hat{\rho}$.  It is worth mentioning that many models considering baryonic mass density do not meet the condition at the boundary radius of the configuration, i.e. $P(R)=P_{b}=0$. We considered these models as non-integrable. Numerical integration was performed in Python, using the \textit{RK45} method with the \textit{solve\_ivp} routine having an accuracy of $10^{-15}$ for vanishing the dimensionless pressure at the surface. 

\subsubsection{Cracking and convective instability}
We wanted to explore the incidence of convective instability with cracking since the second derivative of density appears in the perturbed hydrostatic equilibrium equation for cracking stability (see equation \ref{CrackingConvective}, Appendix \ref{appendix:a}). However, as we can be seen from Figure \ref{figCrackingPlot} when the model is unstable to convective motions (right plate), it does not present cracking within the material configuration. Hence, convective motions do not affect sign change in the perturbed hydrostatic equilibrium equation due to local perturbations.

\subsubsection{Tidal deformability of anisotropic relativistic spheres}
Tidal effects arise on extended bodies when immersed in an external gravitational field and measure the quadrupole deformation in response to a companion perturbating star~\cite{PoissonWill2014}. Several authors studied the influences of anisotropy on the deformability limits of various ultradense EoS (see~\cite{YagiYunez2015, BiswasBose2019, RaposoetEtal2019, RahmansyahSulaksono2021, DasEtal2021, ArbanilPanotopoulos2022} and references therein) through two standard quantities:  the dimensionless tidal polarizability, $\bar{\Lambda}_{\star}$, and Love number $k_{2\star}$ of the each NS. The $\bar{\Lambda}_{\star}$, is often employed in gravitational-wave Astronomy and can be expressed as
\begin{equation}
\bar{\Lambda}_{\star}= \frac{\Lambda_{\star}}{M_{\star}^{5}} = \frac{2 k_{2 \star}}{3 \mathcal{C}_{\star}^{5}}  \, .
\label{LambdaBar}
\end{equation}

The tidal Love number, $k_{2}$, quantifies the deformability of a star~\cite{Hinderer2008, DamourNagar2009, PoissonWill2014} and is calculated in terms of the compactness $\mathcal{C}_{\star}$ and the logarithmic derivative, $y(R)$, of the perturbed metric evaluated at the star's boundary surface, as
\begin{equation}
    \label{lovenumber}
    k_{2}=\frac{A_{1}}{A_{2}} \,,
\end{equation}
where the numerator and denominator are
\begin{equation}
    A_{1} = \frac{8}{5}(1-2 \mathcal{C}_{\star})^{2} \mathcal{C}_{\star}^{5}[2 \mathcal{C}_{\star}(y(R)-1)-y(R)+2] \quad {\rm and}    
\end{equation}
\begin{equation}
    \begin{array}{ll}
    A_{2}& = 2 \mathcal{C}_{\star}\left[4(y(R)+1) \mathcal{C}_{\star}^{4}+(6 y(R)-4) \mathcal{C}_{\star}^{3}\right] +2 \mathcal{C}_{\star}\left[(26-22 y(R)) \mathcal{C}_{\star}^{2}+3(5 y(R)-8) \mathcal{C}_{\star} \right. \\
     &\qquad \left. -3 y(R)+6\right] -3(1-2 \mathcal{C}_{\star})^{2}[2 \mathcal{C}_{\star}(y(R)-1)-y(R)+2] \ln \left(\frac{1}{1-2 \mathcal{C}_{\star}}\right) \, .
    \end{array}
\end{equation}
In the Appendix \ref{appendix:c} we briefly sketch the ideas behind deformability of anisotropic relativistic spheres, and also see references~\cite{RahmansyahSulaksono2021} and ~\cite{DamourNagar2009} for more details. 

After we obtain $k_2$ for each NS with equation (\ref{lovenumber}), and each dimensionless tidal polarizability, $\bar{\Lambda}_{\star}$, from equation (\ref{LambdaBar}), we can calculate mass-weighted tidal deformation, $\tilde{\Lambda}_{(1.4)\star}$, of a particular star with mass, $m_{\star}$ and tidal deformability $\bar{\Lambda}_{\star}$, with respect to companion of mass $m_{(1.4)}=1.4~M_{\odot}$:
\begin{equation}
    \tilde{\Lambda}_{(1.4)\star} = \frac{16}{13} \frac{\left(m_{(1.4)} +12 m_{\star}\right) m_{(1.4)}^{4} \bar{\Lambda}_{(1.4)} 
    +\left(m_{\star}+12 m_{(1.4)}\right) m_{\star}^{4} \bar{\Lambda}_{\star}}{\left(m_{(1.4)} +m_{\star}\right)^{5}} \, .
\end{equation}

From the event GW170817, LIGO reported a constraint on $\tilde{\Lambda}_{(1.4)\star}$  as $50~\leq~\tilde{\Lambda}_{(1.4)\star}~\leq~800$ at the $90\%$ confidence level~\cite{AbbottEtalLIGOVIRGCol2019, ChirentiPosadaGuedes2020}.

\subsection{Neutron stars  candidates}
The models that emerge from the integration of (\ref{TOVAdi})-(\ref{MassAdi}) are physically interesting. In table \ref{tableparameters1} we show four NS that result from numerically integrating equations (\ref{TOVAdi})-(\ref{MassAdi}) using (\ref{TOVAdih}) and (\ref{DelCosenzadi}). The first two could be associated, with respect to their masses, with neutron stars such as the pulsar J0737-30309 $(n=0.5$) of mass $M=1.33 M_\odot$~\cite{BurgayEtal2003, LyneEtal2004} and the pulsar J1518+490 $(n=1.0)$ of mass $M=1.56 M_\odot$~\cite{NiceSayerTaylor1996,JanssenEtal2008}, the estimated radii for these two objects are 11.49 and 9.88 km, respectively. The third compact object, which we will call GMn075 $(n=0.75)$, corresponds to a Generic Model of mass 1.50 $M_\odot$ and radius 10.0 Km. In this paper we will use this generic model to show our results for values of $n=0.5, 1.0, 1.5$ and $n=2.0$, so the above-mentioned objects would be included by the values of $n$ considered. The fourth object is like the third one but with $\alpha=\beta=0$, that is, the polytropic case, already studied in~\cite{HerreraBarreto2004} and which we will call the Polytropic Model PMn075 $(n=0.75)$. 

Figure \ref{FigFracturaMRModels} (right plate) displays the $M-R$ curves for NS candidates considered in Table \ref{tableparameters1}. As we varied the central density, we found a family of stable models for each NS-candidates. Regarding the case of Mass-Radius for the  J1518+4904, we see that it could also represent any of the other candidates.

For all these candidates we compute the dimensionless tidal polarizability, $\bar{\Lambda}_{1.4 \star}$, and Love number $k_{2 \star}$. All the obtained values are less than the critical upper limit estimated by LIGO~\cite{AbbottEtalLIGOVIRGCol2019, ChirentiPosadaGuedes2020}.

In figures \ref{HZNCriterionMass} and \ref{HZNMR} we have indicated the value of $M \approx 2.08 \pm 0.07 M_{\odot}$ corresponding to the recently observed PSR J0740+6620~\cite{MillerEtal2021}. In these plots, it is clear that for the values of the parameters $\alpha$, $\varkappa$ and $n$, considered in our modelling, only anisotropic matter configurations could describe this massive compact object. We have also shown the stiffness threshold of $\sigma =1$, implying that the region with $\sigma \geq 1$ is ruled out for our modelling.

\begin{table}[h!]
\caption{Parameters for the numeric solution of the Master Lane-Emden equations (\ref{TOVAdi})-(\ref{MassAdi}), modelling two NS candidates: the Pulsar J0737-30309, the Pulsar J1518+4904, a Generic Model of the compact star (GMn075) and a Polytropic model (PMn075), which correspond to the case $\alpha=\beta=0$~\cite{SuarezurangoNunezHernandez2021}. The output parameter displayed are: total mass $M$, total radius $R$, NS compactness $\mathcal{C}_{\star}$ at surface, central density $\rho_c$, boundary  density $\rho_b$, the mass-weighted dimensionless tidal polarizability, $\tilde{\Lambda}_{(1.4)*}$, and Love number $k_{2\star}$. }
\centering
\begin{tabular}{|c | r r r r |}
\cline{1-5}
\multicolumn{5}{|c|} {\hspace{2.5 cm} Object} \\ \hline \hline
Input & J0737-3039 & J1518+4904 & GMn075 & PMn075  \\ [-0.5ex]
parameters& $n=0.50$ & $n=1.00$ & $n=0.75$ & $n=0.75$  \\   \hline\hline
$C$ & 0.09 & 0.125 & 0.05 & 0.05    \\ 
$\alpha$ &-0.01 &  0.01  & -0.01 & 0.0    \\
$\varkappa$ & 0.05 &  0.15 & 0.17  & 0.0      \\
$\sigma$ & 0.10 &  0.15  & 0.18 & 0.18      \\
$\rho_c \times 10^{15}$ (g/cm$^3$) & 0.66 &  1.79 & 1.41 & 1.41      \\ [0.5ex] \hline\hline
Output & & & & \\[-0.5ex] 
parameters & & & & \\[0.5ex] \hline\hline
$M$ ($M_\odot$) & 1.33 & 1.56 & 1.50 & 1.56    \\ 
$R$ (km)  &11.49 &  9.88 & 10.0  & 10.9    \\
$2\mathcal{C}_{\star}$  & 0.34 &  0.47  & 0.44 & 0.42     \\
$\rho_b\times 10^{14}$ (g/cm$^3$)& 0.33 &  2.69 & 2.4 & 0.0   \\
$k_{2\star}$ & 0.06 & 0.03 & 0.04 & 0.04 \\
$\tilde{\Lambda}_{(1.4)*}$ & 165.30 & 50.80 & 63.60 & 70.40 \\[0.5ex] 
\hline
\end{tabular} 
\label{tableparameters1}
\end{table}

\section{Conclusions and final remarks}
\label{FinalRemarks}
Modelling compact objects with anisotropic polytropes started in 2013~\cite{HerreraBarreto2013} and generated many exciting candidates. A quick search in ADS retrieves more than fifty papers~\footnote{ADS Database \url{https://ui.adsabs.harvard.edu/} with: Anisotropic pressures AND polytropic Equation of State AND General Relativity}. In this work, we study the EoS introduced previously~\cite{HernandezSuarezurangoNunez2021} and identified the parameter-space portion ensuring the configurations' acceptability.

We refined the acceptability condition {\bf C1} presented by B.V. Ivanov~\cite{Ivanov2017} and later extended in~\cite{HernandezNunezVasquez2018}. We found that the usual restriction on the local compactness, $\mathcal{C}~=~m/r~<~1/2$ is enough to obtain physically reasonable metric coefficients. The Ivanov {\bf C1} acceptability condition on the mass-metric coefficient,   $(m/r)^{\prime} > 0$, should be considered as a sufficient but not a necessary restriction.

We explored compact objects' modelling emerging from the generalization of the polytropic EoS (\ref{MasterPolytropic}). Our framework includes five important physical variables: the polytropic index, $n$; the linear coefficient $\alpha$; the stiffness at the centre of the matter distribution, $\sigma = P_c/\rho_c$; the anisotropic factor $C$ and, $\varkappa =  \rho_b/\rho_c$, the density drop from the centre to the surface of the compact object. This EoS includes several other particular cases found in the literature (see references~\cite{HernandezSuarezurangoNunez2021, NasheehaThirukkaneshRagel2021} and references therein), and the variation of these variables generate a parameter-space representing a wide range of possible astrophysical candidates.  

Following the heuristic approach found in references~\cite{CosenzaEtal1981} and \cite{HerreraBarreto2013}, we included the anisotropic distribution of pressures. This approach is free of the pathologies pointed out in the previous article~\cite{HernandezSuarezurangoNunez2021}. We then integrate the corresponding Lane-Emden structure equations and identify the most significant variables and parameter space zone leading to acceptable astrophysical candidates. 

We implemented several hundred models based on the fulfilment of the nine acceptability conditions, and we found the following answers to our initial questions.
\begin{itemize}
    \item The parameter $n$, $\alpha$, $\sigma$ and $C$ are the most important. The variation of $\varkappa$ does not significantly change the stability of the generated models. 
    \item Low polytropic indexes, $n < 1$, lead to acceptable models and configurations having high stiffness do not comply with the causality condition.   
    \item When the isotropy is perturbed with $P_{\perp} > P$, improves the configurations' acceptability of the models. Small positive anisotropies produce better models than either negative or large anisotropy. The smaller the anisotropy is, the more acceptable models are.
    \item Polytropic matter configurations are more viable when considering the total energy density~\cite{SuarezurangoNunezHernandez2021}. Many models considering baryonic mass density do not meet the condition at the boundary radius of the configuration, i.e. $P(R)=P_{b}=0$.
    \item Unstable models against convective motions do not present cracking. Hence, convective disturbances do not affect the sign change in the perturbed hydrostatic equilibrium equation due to local perturbations.
    \item The models emerging from our simulations could represent physically interesting astrophysical objects. In table \ref{tableparameters1} we show four of these NS candidates. The  mass-weighted tidal polarizability, $\tilde{\Lambda}_{(1.4)\star}$ for all these models are less than the critical upper limit estimated by LIGO~\cite{AbbottEtalLIGOVIRGCol2019, ChirentiPosadaGuedes2020}.
    \item Regarding the massive recently observed pulsar J0740+6620~\cite{MillerEtal2021} we found that, for the values of the parameters $\alpha$, $\varkappa$ and $n$ considered in our modelling, only anisotropic matter configurations could describe this massive compact object. 
\end{itemize}

These findings are not general but depend on the provided heuristic relation between radial and tangential pressure given in equation (\ref{ansatzcosenza}) for the generalized polytropic EoS (\ref{MasterPolytropic}). 

As we have mentioned, there are diverse heuristic strategies to generate an anisotropic distribution of pressures within matter distributions: the standard one --discussed in our previous work~\cite{HernandezSuarezurangoNunez2021}--; the present approach described in equation (\ref{DelCosenza}) ~\cite{CosenzaEtal1981, HerreraBarreto2013};  the original one~\cite{BowersLiang1974}; the quasilocal assumption~\cite{DonevaYazadjiev2012}; the covariant way~\cite{RaposoEtal2019}; Karmarkar embedding class I~\cite{Karmarkar1948, OspinoNunez2020}; conditions on the complexity factor~\cite{Herrera2018};   double polytrope schemes~\cite{AbellanEtal2020B} and gravitational decoupling~\cite{AbellanEtal2020C}. These strategies may lead to viable interesting astrophysical  models~\cite{Setiawan2019}  within particular acceptability parameter spaces. We are exploring how similar the above answers are for all these other strategies. This is a work in progress and will be reported shortly.

\section*{Acknowledgments}
We want to thank the comments and criticism from the anonymous referees, which lead us to the significant improvement of the final version of our manuscript. The authors also thank Prof. Jos\'e Natario for pointing us some inaccurate statements in our manuscript, and Jes\'us Pe\~{n}a-Rodr\'iguez for the fruitful discussion and ideas contributed to the visualization of the results.  We gratefully acknowledge the financial support of the Vicerrector\'ia de Investigaci\'on y Extensi\'on de la Universidad Industrial de Santander and the financial support provided by COLCIENCIAS under Grant No. 8863. J.O. express gratitude for financial support from Spain Ministerio de Ciencia, Innovaci\'on y Universidades, Grant number: PGC2018-096038-B-I00, and Junta de Castilla y Le\'on, Grant number: SA096P20.

\section*{Appendices}
\appendix

\section{On the criteria of cracking and adiabatic convectivity}
\label{appendix:a}
Just for completeness we shall consider in this Appendix the local perturbations of density, $\delta \rho~=~\delta \rho(r)$, and show the difference between the present  {\bf C8} and the previous more simple cracking criterion~\cite{AbreuHernandezNunez2007b}. The $\delta \rho(r)$ fluctuations induce variations in all the other physical variables, i.e. $m(r), P(r), P_\perp(r)$ and their derivatives, generating a non-vanishing total radial force distribution. For further details, we refer interested readers to~\cite{HernandezSuarezurangoNunez2021,HernandezNunezVasquez2018, GonzalezNavarroNunez2015,GonzalezNavarroNunez2017} and references therein. 

Following~\cite{GonzalezNavarroNunez2017}, we take equation (\ref{TOVStructure1}) and define:
\begin{equation}
 \label{Ranitov}
 \mathcal{R} \equiv \frac{\mathrm{d} P}{\mathrm{d} r} +(\rho +P)\frac{m + 4 \pi r^{3}P}{r(r-2m)} -\frac{2}{r}\left(P_\perp-P \right) \, .
\end{equation}
Next, expanding this TOV equation as $\mathcal{R} \approx \mathcal{R}_{0}(\rho, P, P_\perp, m, P^{\prime}) + \delta \mathcal{R}$, thus 
\begin{equation}
\label{RAniExpanded}
\delta \mathcal{R} \equiv 
\frac{\partial \mathcal{R}}{ \partial \rho} \delta \rho
+\frac{\partial \mathcal{R}}{ \partial P} \delta P
+ \frac{\partial \mathcal{R}}{\partial P_\perp} \delta P_\perp
+\frac{\partial \mathcal{R}}{ \partial m} \delta m 
+\frac{\partial \mathcal{R}}{ \partial P^{\prime}} \delta P' \,,
\end{equation}
where $\mathcal{R}_{0}(\rho, P, P_\perp, m, P^{\prime})=0$, because initially the configuration is in equilibrium.

The above equation (\ref{RAniExpanded}) can be reshaped as:
\begin{equation}
 \delta \mathcal{R} \equiv \delta \underbrace{P'}_{\mathcal{R}_p}  + \delta \underbrace{\left[ (\rho +P)\frac{m + 4 \pi r^{3}P}{r(r-2m)} \right]}_{\mathcal{R}_g}  + \delta \underbrace{\left( 2\frac{P}{r}- 2\frac{P_\perp}{r}\right)}_{\mathcal{R}_a} = \delta \mathcal{R}_p + \delta \mathcal{R}_g +\delta \mathcal{R}_a \, ,
 \label{PartdeltaR}
\end{equation}
where it is clear that the density perturbations $\delta \rho(r)$ are influence the distribution of reacting pressure forces $\mathcal{R}_p$, gravity forces $\mathcal{R}_g$ and anisotropy forces $\mathcal{R}_a$.  Depending on this effect, each perturbed distribution force can contribute in a different way to the change of sign of $\delta \mathcal{R}$: each term can be written as 
\begin{equation}
\label{Fp_pg}
\delta \mathcal{R}_p = \left( \frac{P''}{\rho'} \right) \delta \rho = \left((v_s^2)' + v_s^2\frac{\rho''}{\rho'}\right) \delta \rho \, , \quad
\delta \mathcal{R}_g = \left( \frac{ \partial \mathcal{R}_g }{ \partial \rho } + \frac{ \partial \mathcal{R}_g }{ \partial P } v_s^2 + \frac{ \partial \mathcal{R}_g }{ \partial m } \frac{4 \pi r^2 \rho}{\rho'}\right) \delta \rho \quad \textrm{and}
\end{equation}
\begin{equation}
\label{PerturbAnisotropy}
\delta \mathcal{R}_a = \left( \frac{ v_s^2 - v_{s_\perp}^2 }{ r } \right) \delta \rho ,
\end{equation}
with
\begin{equation}
\label{Fg_rho}
\frac{\partial \mathcal{R}_g}{\partial \rho}  =  \frac{m+  4 \pi r^3 P }{ r(r - 2m)}, \quad
\frac{\partial \mathcal{R}_g}{\partial P} = \left[ \frac{ m + 4 \pi r^3 ( \rho + 2 P )}{r(r-2m)} \right]
\; \; \textrm{and} \; \;
\frac{\partial \mathcal{R}_g}{\partial m}  =  \left[\frac{ (\rho + P)( 1 + 8\pi r^2 P)  }{(  2 m  - r )^2 }\right].
\end{equation}
Notice that if, as in~\cite{AbreuHernandezNunez2007b}, the perturbation $\delta \rho$ is constant and does not affect the pressure gradient, we have: $\delta \mathcal{R}_p = 0$, 
\begin{equation}
\label{AbreuEtAl}
\delta \tilde{\mathcal{R}_g} = \left(2\frac{ m + 4 \pi r^3 ( \rho + 2 P )}{r(r-2m)} +\frac{4 \pi r^2}{3} \frac{ (\rho + P)( 1 + 8\pi r^2 P)  }{(  2 m  - r )^2}\right)\delta \rho, \quad \textrm{and } \;
\delta \tilde{\mathcal{R}_a} = \left( \frac{ v_s^2 - v^2_{s_\perp}}{ r } \right) \,.
\end{equation}
Thus, only anisotropic matter distribution can present cracking instabilities because $\delta \tilde{\mathcal{R}_g} >0$ for all $r$ and the possible change of sign for $ \delta \mathcal{R}$ should emerge from $\delta \mathcal{R}_a$ and the criterion against cracking is written as: 
\begin{equation}
\label{CrackingCriterion}
-1 \leq v_{s_\perp}^2 -v_s^2 \leq 0 \quad \Leftrightarrow \quad 
0 \geq \frac{\mathrm{d} P_\perp}{\mathrm{d} r} \geq \frac{\mathrm{d} P}{\mathrm{d} r} \,.
\end{equation}

For the present anisotropic case $\mathcal{R}$ is
\begin{equation}
    \mathcal{R} \equiv \frac{\mathrm{d}P}{\mathrm{d}r} + h(\rho + P) \frac{m + 4 \pi r^{3} P}{r(r-2m)} = 0 \, ,
\end{equation}
and  we get
\begin{eqnarray}
 \frac{\partial \mathcal{R}}{\partial \rho} &=& \frac{h \left(m + 4 \pi r^{3} P \right)}{r(r-2m)} \,, \qquad  \quad 
 \frac{\partial \mathcal{R}}{\partial m} =  \frac{h \left(\rho + P \right) \left(1 + 8 \pi P r^{2} \right)}{\left(r - 2m \right)^{2}} \, , \nonumber \\
 \frac{\partial \mathcal{R}}{\partial P} &=&  \frac{h \left[m + 4 \pi r^{3} \left(\rho + 2 P \right) \right]}{r(r-2m)}  \quad \mbox{and} \quad 
  \frac{\partial \mathcal{R}}{\partial P^{\prime}} = 1 \,. \nonumber
\end{eqnarray}

Hence, we have
\begin{eqnarray}
    \frac{\delta \mathcal{R}}{\delta \rho} &=  &\frac{h \left(m + 4 \pi r^{3} P \right)}{r(r-2m)}  + \frac{h \left(\rho + P \right) \left(1 + 8 \pi P r^{2} \right)}{\left(r - 2m \right)^{2}} \, \frac{4 \pi r^{2} \rho}{\rho^{\prime}} \nonumber \\ &&+  \frac{h \left[m + 4 \pi r^{3} \left(\rho + 2 P \right) \right]}{r(r-2m)} v^{2}  + \left[\left(v^{2} \right)^{\prime} + v^{2} \frac{\rho^{\prime \prime}}{\rho^{\prime}} \right] \, ,
    \label{CrackingConvective}
\end{eqnarray}
and in the Lane-Emden variables we get
\begin{eqnarray}
\bar{\mathcal{R}}= a \frac{\delta \mathcal{R}}{\delta \rho} &\equiv &\frac{h\Upsilon (1+n)}{n} \left[\frac{ n\left[\eta + \xi^{3} \mathcal{P} \right] }{\xi \left[\xi - 2 \Upsilon \left(1+n \right) \eta \right]} +  \frac{  \xi^{2} \Psi \left[\Psi^{n} + \mathcal{P} \right][1 + 2 \Upsilon \left(1+n \right)  \mathcal{P}  \xi^{2}]}{ \dot{\Psi} \left[\xi - 2 \Upsilon \left(1+n \right) \eta \right]^{2}} \right. \nonumber \\
 &+& \left.  \frac{1}{\Xi} \frac{ \Upsilon \left(1+n \right) \left[\eta + \xi^{3} \left[\Psi^{n} + 2  \mathcal{P} \right] \right] \Psi }{ \xi \left[\xi - 2 \Upsilon \left(n+1 \right) \eta \right]} + \frac{n \dot{\Psi}^{2} + \Psi \ddot{\Psi}  }{h  \dot{\Psi}}  \right] + \alpha \left[\frac{(n-1)\dot{\Psi}}{\Psi}+ \frac{\ddot{\Psi}}{\dot{\Psi}}\right] \, , \end{eqnarray}
 where
 \begin{equation*}
    \Xi=\left[1+\frac{\alpha n}{\Upsilon(1+n) \Psi}\right]^{-1} \,.
    \label{Aux}
\end{equation*}
 
On the other hand, in reference~\cite{HernandezNunezVasquez2018} is developed a simple criterion to identify unstable state equations to convection and explore the influence of buoyancy on cracking (or overturning) for isotropic and anisotropic relativistic spheres. The criterion of adiabatic stability against convection consists of analyzing a fluid element displaced towards the sphere's centre and its interaction with the surrounding environment. It is found that the material configuration will be stable under this type of disturbance if the second derivative of the density concerning the radius is less than or equal to zero ($\rho''\leq 0$).

Starting from the density equation
\begin{equation*}
      \rho = \rho_c \Psi^n(\xi) \,,
\end{equation*}
we compute the second derivative of density respect to $\xi$ as follows
\begin{equation}
     {\ddot \Psi^{n}} = n \frac{\Psi^{n}}{\Psi}\left[ \left(n-1 \right) \frac{\dot{\Psi}^{2}}{\Psi} + \ddot{\Psi} \right] \,.
\end{equation}

\section{The ``master'' Lane-Emden equation}
\label{appendix:b}
The hydrostatic equilibrium equation (\ref{ansatzcosenza}) can be written as
\begin{equation}
\label{EEHA}
    \frac{r \left(r- 2m \right)}{\rho + P} \frac{\mathrm{d}P}{\mathrm{d}r} + h \left(m + 4 \pi P r^{3} \right) = 0 \,.
\end{equation}

Using a change of variables for energy density, $\rho = \rho_{c} \Psi^{n}$, we have
\begin{equation}
\label{PresionLEMA}
    P = K \rho_{c}^{\gamma} \Psi^{n \gamma} + \alpha \rho_{c} \Psi^{n} - \beta = K \rho_{c}^{1 + \frac{1}{n}} \Psi^{n+1} + \alpha \rho_{c} \Psi^{n} - \beta
\end{equation}
and
\begin{equation}
\label{GradientePresionLEMA}
    \frac{\mathrm{d}P}{\mathrm{d}r} = K \rho_{c}^{1 + \frac{1}{n}} (n+1) \Psi^{n} \frac{\mathrm{d}\Psi}{\mathrm{d}r} + \alpha \rho_{c} n \Psi^{n-1} \frac{\mathrm{d}\Psi}{\mathrm{d}r}\,.  
\end{equation}

Substituting the energy density, (\ref{PresionLEMA}) and (\ref{GradientePresionLEMA}) in (\ref{EEHA}), yields

\begin{equation}
    \frac{r \left(r -2m \right)}{1 + K \rho_{c}^{\frac{1}{n}} \Psi + \alpha - \beta / \rho_{c} \Psi^{n}  } \left[K \rho_{c}^{\frac{1}{n}} (n+1) + \frac{\alpha n}{\Psi} \right] \frac{\mathrm{d}\Psi}{\mathrm{d}r}
    + h \left[m + 4 \pi \left(K \rho_{c}^{1 + \frac{1}{n}} \Psi^{n+1} + \alpha \rho_{c} \Psi^{n} - \beta \right) r^{3} \right] = 0.
\end{equation}

Introducing the radial coordinate as $r = a \xi$, where
\begin{equation*}
    a^{2} = \frac{\Upsilon \left(1+n\right)}{4 \pi \rho_{c}} \quad \textrm{with} \quad  \Upsilon = \kappa \rho_c^{1/n} = \frac{\sigma - \alpha \left(1 - \varkappa \right)}{1 - \varkappa^{1 + 1/n}} ; \quad \varkappa = \frac{\rho_{b}}{\rho_{c}}\,,
\end{equation*}
we have
\begin{equation}
    \frac{a^{2} \xi \left[\xi -  2m/a  \right]}{\left(1 + \alpha \right) + \Upsilon \Psi  - \left(\beta / \rho_{c} \right) \Psi^{-n}} \left[\Upsilon (n+1) + \frac{\alpha n}{\Psi} \right] \frac{1}{4 \pi \rho_{c} a^{4}} \frac{\mathrm{d}\Psi}{\mathrm{d}\xi} \\
    + h \left\{\frac{m}{4 \pi \rho_{c} a^{3}} + \xi^{3} \left[\Upsilon \Psi^{n+1} + \alpha \Psi^{n} - \beta/\rho_{c}  \right] \right\} = 0.
\end{equation}

Now, the dimensionless mass is
\begin{equation}
    \eta (\xi) = \frac{m}{4 \pi a^{3} \rho_{c}}  \,,
\end{equation}
yielding
\begin{equation}
    \frac{\xi \left[\xi - 2\left(4\pi \rho_{c} a^{2} \eta  \right)  \right]}{\left(1 + \alpha \right) + \Upsilon \Psi  - \left(\beta / \rho_{c} \right) \Psi^{-n}} \left[\Upsilon (n+1) + \frac{\alpha n}{\Psi} \right] \frac{1}{4 \pi \rho_{c} a^{2}} \frac{\mathrm{d}\Psi}{\mathrm{d}\xi} \\
    + h \left\{\eta + \xi^{3} \left[\Upsilon \Psi^{n+1} + \alpha \Psi^{n} - \beta/\rho_{c}  \right] \right\} = 0 \,.
\end{equation}

Finally, making
\begin{equation*}
    \frac{\Upsilon(n+1)}{4 \pi a^{2}\rho_{c}} = 1 \,,
\end{equation*}
we get
\begin{equation}
\label{EEHLEA2}
    \frac{\xi \left[\xi - 2 \Upsilon \left(n+1 \right) \eta  \right]}{\left(1 + \alpha \right) + \Upsilon \Psi  - \left(\beta / \rho_{c} \right) \Psi^{-n}} \left[1 + \frac{\alpha n}{\Upsilon (n+1) \Psi} \right] \frac{\mathrm{d}\Psi}{\mathrm{d}\xi} \\
    + h \left\{\eta + \xi^{3} \left[\Upsilon \Psi^{n+1} + \alpha \Psi^{n} - \beta/\rho_{c}  \right] \right\} = 0 \,.
\end{equation}

By using (\ref{betamasters}),  equation (\ref{EEHLEA2}) becomes
\begin{equation}
\label{EEHLEA3}
    \frac{\xi \left[\xi - 2 \Upsilon \left(n+1 \right) \eta  \right]}{\left(1 + \alpha \right) + \Upsilon \Psi  - \left(\beta / \rho_{c} \right) \Psi^{-n}} \left[1 + \frac{\alpha n}{\Upsilon (n+1) \Psi} \right] \frac{\mathrm{d}\Psi}{\mathrm{d}\xi} \\
    + h \left\{\eta + \xi^{3} \left[\Upsilon \left( \Psi^{n+1} - \varkappa^{1 + \frac{1}{n}} \right) + \alpha \left(\Psi^{n} - \varkappa  \right) \right] \right\} = 0 \,,
\end{equation}
and together with equation (\ref{MassAdi}) they form master Lane-Emden equation for the generalized polytropic equation of state.

\section{Tidal deformability of anisotropic relativistic spheres}
\label{appendix:c}
When a companion NS provides an external gravitational tidal field ($E_{ij}$) generating a quadrupole moment ($Q_{i j}$) on the corresponding NS. This moment is 
\begin{equation}
Q_{i j}=-\frac{2 k_{2\star} R^{5}}{3} E_{i j} \equiv-\Lambda_{\star} E_{i j}\, .
\label{quadrupole}
\end{equation}
Here, $k_{2\star}$ is the tidal Love number quantifying the deformability of a star immersed in an external field~\cite{PoissonWill2014} and $\bar{\Lambda}_{\star}$, the dimensionless tidal polarizability often employed in gravitational-wave astronomy. It can be expressed as
\begin{equation}
\bar{\Lambda}_{\star}= \frac{\Lambda_{\star}}{M_{\star}^{5}} = \frac{2 k_{2 \star}}{3 \mathcal{C}_{\star}^{5}}  \, ,
\label{LambdaBarApp}
\end{equation}
where $\mathcal{C}_{\star} \equiv M_{\star}/R_{\star}$  denotes the compactness of the matter configuration.

Following~\cite{RahmansyahSulaksono2021} and perturbing the metric (\ref{metricSpherical}) and its corresponding field equations (\ref{FErho}), (\ref{FEPrad}) and (\ref{FEPtan}), we obtain a second order differential equation for the even parity metric perturbations $H(r)$, 
\begin{equation}
   H^{\prime \prime} +D_{1}(r) H^{\prime}  +D_{0}(r) H=0 \, ,
   \label{Hequation}
\end{equation}
where the coefficients $D_1(r)$ and $D_0(r)$ are given by
\begin{equation}
D_{1}(r)= \frac{2}{r}+e^{2 \lambda}\left(\frac{2 m}{r^{2}}+4 \pi r(P-\rho)\right)  \quad   {\rm and}
\label{CoeffD1}
\end{equation}
\begin{equation}
D_{0}(r)= e^{2 \lambda}
\left( -\frac{6}{r^{2}} +\frac{4 \pi(P +\rho)(1+v^2_{s})}{v^2_{s \perp}} +4 \pi(4 \rho +8 P)\right)  
+16 \pi ( P -P_{\perp}) e^{2 \lambda} -  \left(\nu^{\prime} \right)^{2} \, ,    
\label{CoeffD0}
\end{equation}
see references~\cite{DamourNagar2009} and~\cite{RahmansyahSulaksono2021} for details.  

Next, to simplify the integration of the perturbation equation (\ref{Hequation}), it is conventional to introduce the logarithmic derivative $y(r)\equiv rH^{\prime}/H$, to obtain a Riccati equation~\cite{DamourNagar2009}, 
\begin{equation}
  r y^{\prime} +y(y-1) +r D_1 y + r^2 D_0 = 0 \, .
  \label{Ricattiperturb}
\end{equation}

Now, integrating the system (\ref{TOVAdi}), (\ref{MassAdi}) \& (\ref{Ricattiperturb}), guaranteeing the continuity of $y$ and its derivative across the boundary surface, $r = R$
\begin{equation}
    \left. y \right|_{R_{-}} = \left. y \right|_{R_{+}} = y(R) \quad {\rm and} \quad
    \left. y^{\prime} \right|_{R_{-}} = \left. y^{\prime} \right|_{R_{+}} = y^{\prime}(R) \,,
\end{equation}
it is possible to calculate the Love number $k_2$ from the following expression
\begin{equation}
    k_{2}=\frac{A_{1}}{A_{2}}\,,
\end{equation}
where
\begin{equation}
    A_{1} = \frac{8}{5}(1-2 \mathcal{C}_{\star})^{2} \mathcal{C}_{\star}^{5}[2 \mathcal{C}_{\star}(y(R)-1)-y(R)+2], \quad {\rm and}    
\end{equation}
\begin{equation}
    \begin{array}{ll}
    A_{2}& = 2 \mathcal{C}_{\star}\left[4(y(R)+1) \mathcal{C}_{\star}^{4}+(6 y(R)-4) \mathcal{C}_{\star}^{3}\right] +2 \mathcal{C}_{\star}\left[(26-22 y(R)) \mathcal{C}_{\star}^{2}+3(5 y(R)-8) \mathcal{C}_{\star} \right. \\
     &\qquad \left. -3 y(R)+6\right] -3(1-2 \mathcal{C}_{\star})^{2}[2 \mathcal{C}_{\star}(y(R)-1)-y(R)+2] \ln \left(\frac{1}{1-2 \mathcal{C}_{\star}}\right) \,,
    \end{array}
\end{equation}
see references~\cite{DamourNagar2009} and~\cite{RahmansyahSulaksono2021} for details. 

After we obtain $k_2$ for each NS, we can calculate mass-weighted combination of a star with mass, $m_{\star}$ and tidal deformability $\Lambda_{\star}$, with respect to companion star of mass $m_{(1.4)}=1.4~M_{\odot}$: 
\begin{equation}
    \tilde{\Lambda}_{(1.4)}=\frac{16}{13} \frac{\left(m_{(1.4)} +12 m_{\star}\right) m_{(1.4)}^{4} \bar{\Lambda}_{(1.4)} 
    +\left(m_{\star}+12 m_{(1.4)}\right) m_{\star}^{4} \bar{\Lambda}_{\star}}{\left(m_{(1.4)} +m_{\star}\right)^{5}} \,,
\end{equation}
that can be compared with data from GW observations.


\begin{thebibliography}{10}

\bibitem{Wolansky1999}
G.~Wolansky.
\newblock On nonlinear stability of polytropic galaxies.
\newblock {\em ANN I H POINCARE-AN},
  16(1):15--48, 1999.

\bibitem{Chandrasekhar1967}
S.~{Chandrasekhar}.
\newblock {\em {An introduction to the study of stellar structure}}.
\newblock Dover, New York, 1967.

\bibitem{Tooper1964}
R.~F. Tooper.
\newblock General relativistic polytropic fluid spheres.
\newblock {\em Astrophys. J.}, 140:434 -- 459, 1964.

\bibitem{Kovetz1967}
A.~Kovetz.
\newblock Schwarzschild's criterion for convective instability in general
  relativity.
\newblock {\em Zeitschrift f{\"u}r Astrophysik}, 66:446, 1967.

\bibitem{Fowler1964}
W.~A. Fowler.
\newblock Massive stars, relativistic polytropes, and gravitational radiation.
\newblock {\em Rev. Mod. Phys.}, 36(2):545--555, 1964.

\bibitem{RappaportVerbunt1983}
S.~{Rappaport}, F.~{Verbunt}, and P.~C. {Joss}.
\newblock A new technique for calculations of binary stellar evolution
  application to magnetic braking.
\newblock {\em Astrophys. J.}, 275:713--731, 1983.

\bibitem{CookShapiroTeukolsky1994}
G.~B. {Cook}, S.~L. {Shapiro}, and S.~A. {Teukolsky}.
\newblock Rapidly rotating polytropes in general relativity.
\newblock {\em Astrophys. J.}, 422:227--242, 1994.

\bibitem{Jeans1922}
J.~H. Jeans.
\newblock {The motions of stars in a Kapteyn universe}.
\newblock {\em Mon. Not. R. Astron. Soc.}, 82:122--132, 1922.

\bibitem{Lemaitre1933}
G.~Lema{\i}tre.
\newblock {L'univers en expansion}.
\newblock {\em Ann. Soc. Sci.(Bruxelles) A}, 53:51--85, 1933.

\bibitem{Ruderman1972}
M.~Ruderman.
\newblock Pulsars: Structure and dynamics.
\newblock {\em Annu. Rev. Astron. Astrophys.}, 10:427--476, 1972.

\bibitem{BowersLiang1974}
R.~L. Bowers and E.~P.~T. Liang.
\newblock Anisotropic spheres in general relativity.
\newblock {\em Astrophys. J.}, 188:657--665, 1974.

\bibitem{CosenzaEtal1981}
M.~Cosenza, L.~Herrera, M.~Esculpi, and L.~Witten.
\newblock Some models of anisotropic spheres in general relativity.
\newblock {\em J. Math. Phys.}, 22:118, 1981.

\bibitem{HerreraNunez1989}
L.~{Herrera} and L.~{N\'u{\~n}ez}.
\newblock {Modeling ``hydrodynamic phase transitions'' in a radiating
  spherically symmetric distribution of matter}.
\newblock {\em Astrophys. J.}, 339:339--353, April 1989.

\bibitem{HerreraSantos1997}
L.~Herrera and N.~O. Santos.
\newblock {Local anisotropy in self-gravitating systems}.
\newblock {\em Phys. Rep.}, 286(2):53--130, 1997.

\bibitem{MartinezRojasCuesta2003}
A.~P. Mart\'inez, H.~P. Rojas, and H.~M. Cuesta.
\newblock Magnetic collapse of a neutron gas: Can magnetars indeed be formed?
\newblock {\em Eur. Phys. J. C}, 29(1):111--123, 2003.

\bibitem{HerreraBarreto2004}
L.~{Herrera} and W.~{Barreto}.
\newblock Evolution of relativistic polytropes in the post-quasi-static regime.
\newblock {\em Gen. Relativ. Gravitation}, 36(1):127--150, 2004.

\bibitem{HerreraEtal2014}
L.~Herrera et~al.
\newblock Dissipative collapse of axially symmetric, general relativistic
  sources: a general framework and some applications.
\newblock {\em Phys. Rev. D}, 89(8):084034, 2014.

\bibitem{Setiawan2019}
A.~M. Setiawan and A.~Sulaksono.
\newblock {Anisotropic neutron stars and perfect fluid's energy conditions}.
\newblock {\em Eur. Phys. J. C}, 79(9):755, 2019.

\bibitem{Herrera2020}
L.~Herrera.
\newblock Stability of the isotropic pressure condition.
\newblock {\em Phys. Rev. D}, 101(10):104024, 2020.

\bibitem{Ivanov2017}
B.~V. {Ivanov}.
\newblock {Analytical study of anisotropic compact star models}.
\newblock {\em Eur. Phys. J. C}, 77(11):738, 2017.

\bibitem{Ivanov2018}
B.~V. Ivanov.
\newblock {A conformally flat realistic anisotropic model for a compact star}.
\newblock {\em Eur. Phys. J. C}, 78(4):332, 2018.

\bibitem{DonevaYazadjiev2012}
D.~D. Doneva and S.~S. Yazadjiev.
\newblock Nonradial oscillations of anisotropic neutron stars in the cowling
  approximation.
\newblock {\em Phys. Rev. D}, 85(12):124023, 2012.

\bibitem{RaposoEtal2019}
G.~Raposo et~al.
\newblock Anisotropic stars as ultracompact objects in general relativity.
\newblock {\em Phys. Rev. D}, 99(10):104072, 2019.

\bibitem{Karmarkar1948}
K.~R. Karmarkar.
\newblock Gravitational metrics of spherical symmetry and class one.
\newblock {\em  Proc. Indian Acad. Sci.},
  27(1):56, 1948.

\bibitem{OspinoNunez2020}
J.~{Ospino} and L.~A. {N\'u\~nez}.
\newblock Karmarkar scalar condition.
\newblock {\em Eur. Phys. J. C}, page 166, January 2020.

\bibitem{Ovalle2017}
J.~Ovalle.
\newblock Decoupling gravitational sources in general relativity: from perfect
  to anisotropic fluids.
\newblock {\em Phys. Rev. D}, 95(10):104019, 2017.

\bibitem{AbellanEtal2020C}
G.~Abell{\'a}n, {\'A}.~Rinc{\'o}n, E.~Fuenmayor, and E.~Contreras.
\newblock Anisotropic interior solution by gravitational decoupling based on a
  non-standard anisotropy.
\newblock {\em Eur. Phys. J. Plus}, 135(7):606, 2020.

\bibitem{AbellanEtal2020B}
G.~Abell{\'a}n, E.~Fuenmayor, E.~Contreras, and L.~Herrera.
\newblock The general relativistic double polytrope for anisotropic matter.
\newblock {\em Phys. Dark Universe}, 30:100632, 2020.

\bibitem{Herrera2018}
L.~Herrera.
\newblock New definition of complexity for self-gravitating fluid
  distributions: The spherically symmetric, static case.
\newblock {\em Phys. Rev. D}, 97(4):044010, 2018.

\bibitem{Stewart1982}
B.~W. Stewart.
\newblock Conformally flat, anisotropic spheres in general relativity.
\newblock {\em J. Phys. A: Math. Gen.}, 15(8):2419--2427, 1982.

\bibitem{FinchSkea1989}
M.~R. {Finch} and J.~E.~F. {Skea}.
\newblock {A realistic stellar model based on an ansatz of Duorah and Ray}.
\newblock {\em Class. Quantum Grav.}, 6(4):467--476, 1989.

\bibitem{HernandezNunez2004}
H.~Hern{\'a}ndez and L.~A. N{\'u}{\~n}ez.
\newblock Nonlocal equation of state in anisotropic static fluid spheres in
  general relativity.
\newblock {\em Can. J. Phys.}, 82(1):29--51, 2004.

\bibitem{HerreraOspinoDiPrisco2008}
L.~Herrera, J.~Ospino, and A.~Di~Prisco.
\newblock {All static spherically symmetric anisotropic solutions of Einstein's
  equations}.
\newblock {\em Phys. Rev. D}, 77(2):027502, 2008.

\bibitem{HernandezNunez2013}
H.~Hern{\'a}ndez and L.~A. N{\'u}{\~n}ez.
\newblock {Plausible families of compact objects with a nonlocal equation of
  state}.
\newblock {\em Can. J. Phys.}, 91(4):328--336, 2013.

\bibitem{AbellanEtal2020}
G.~Abell{\'a}n, P.~Bargue{\~n}o, E.~Contreras, and E.~Fuenmayor.
\newblock All static spherically symmetric anisotropic solutions for general
  relativistic polytropes.
\newblock {\em Int. J. Mod. Phys. D}, 29(12):2050082,
  2020.

\bibitem{NasheehaThirukkaneshRagel2021}
R.N. Nasheeha, S.~Thirukkanesh, and F.C. Ragel.
\newblock Anisotropic models for compact star with various equation of state.
\newblock {\em Eur. Phys. J. Plus}, 136(1):1--20, 2021.

\bibitem{HernandezSuarezurangoNunez2021}
H.~Hern{\'a}ndez, D.~Su{\'a}rez-Urango, and L.A. N{\'u}{\~n}ez.
\newblock Acceptability conditions and relativistic barotropic equation of
  state.
\newblock {\em Eur. Phys. J. C}, 81(241), 2021.

\bibitem{BiswasBose2019}
B.~Biswas and S.~Bose.
\newblock Tidal deformability of an anisotropic compact star: Implications of
  gw170817.
\newblock {\em Phys. Rev. D}, 99(10):104002, 2019.

\bibitem{RahmansyahSulaksono2021}
A.~Rahmansyah and A.~Sulaksono.
\newblock Recent multimessenger constraints and the anisotropic neutron star.
\newblock {\em Phys. Rev. C}, 104(6):065805, 2021.

\bibitem{ThirukkaneshRagel2012}
S.~{Thirukkanesh} and F.~C. {Ragel}.
\newblock {Exact anisotropic sphere with polytropic equation of state}.
\newblock {\em Pramana}, 78(5):687--696, 2012.

\bibitem{Ngubelanga2015}
S.~A. Ngubelanga, S.~D. Maharaj, and S.~Ray.
\newblock Compact stars with quadratic equation of state.
\newblock {\em Astrophys. Space Sci.}, 357(1):74, 2015.

\bibitem{TakisaMaharaj2013}
P.~M. {Takisa} and S.~D. {Maharaj}.
\newblock {Some charged polytropic models}.
\newblock {\em Gen. Relativ. Gravitation}, 45(10):1951--1969, 2013.

\bibitem{Malaver2015}
M.~Malaver.
\newblock {Polytropic stars with Tolman IV type potential}.
\newblock {\em AASCIT Journal of Physics}, 1(4):309--314, 2015.

\bibitem{NgubelangaMaharaj2017}
S.~A. Ngubelanga and S.~D. Maharaj.
\newblock New classes of polytropic models.
\newblock {\em Astrophys. Space Sci.}, 362(3):43, 2017.

\bibitem{SharifSadiq2018}
M.~Sharif and S.~Sadiq.
\newblock Cracking in anisotropic polytropic models.
\newblock {\em Mod. Phys. Lett. A}, 33(24):1850139, 2018.

\bibitem{AzamNazir2020}
M.~Azam and I.~Nazir.
\newblock Cracking of some polytropic models via local density perturbations.
\newblock {\em Can. J. Phys.}, 0(ja), 2020.

\bibitem{AbbotEtalLigoCol2017}
B.P. Abbott, R.~Abbott, and T.D.~Abbott et~al.
\newblock {Multi-messenger Observations of a Binary Neutron Star Merger}.
\newblock {\em Astrophys. J. Lett.}, 848(2):L12, 2017.

\bibitem{AbbottEtalLIGOVIRGCol2019}
B.~P. Abbott, R.~Abbott, T.~D. Abbott, et~al, LIGO~Scientific Collaboration,
  and Virgo Collaboration.
\newblock Properties of the binary neutron star merger gw170817.
\newblock {\em Phys. Rev. X}, 9:011001, Jan 2019.

\bibitem{MillerEtal2019}
M.~C. {Miller}, F.~K. {Lamb}, A.~J. {Dittmann}, S.~{Bogdanov},
  Z.~{Arzoumanian}, K.~C. {Gendreau}, S.~{Guillot}, A.~K. {Harding}, W.~C.~G.
  {Ho}, J.~M. {Lattimer}, R.~M. {Ludlam}, S.~{Mahmoodifar}, S.~M. {Morsink},
  P.~S. {Ray}, T.~E. {Strohmayer}, K.~S. {Wood}, T.~{Enoto}, R.~{Foster},
  T.~{Okajima}, G.~{Prigozhin}, and Y.~{Soong}.
\newblock {PSR J0030+0451 Mass and Radius from NICER Data and Implications for
  the Properties of Neutron Star Matter}.
\newblock {\em Astrophys. J. Lett.}, 887(1):L24, December 2019.

\bibitem{MillerEtal2021}
M.~C. Miller, F.~K. Lamb, A.~J. Dittmann, S.~Bogdanov, Z.~Arzoumanian, K.~C.
  Gendreau, S.~Guillot, W.~C.~G. Ho, J.~M. Lattimer, M.~Loewenstein, S.~M.
  Morsink, P.~S. Ray, M.~T. Wolff, C.~L. Baker, T.~Cazeau, S.~Manthripragada,
  C.~B. Markwardt, T.~Okajima, S.~Pollard, I.~Cognard, H.~T. Cromartie,
  E.~Fonseca, L.~Guillemot, M.~Kerr, A.~Parthasarathy, T.~T. Pennucci,
  S.~Ransom, and I.~Stairs.
\newblock The radius of {PSR} j0740+6620 from {NICER} and {XMM}-newton data.
\newblock {\em Astrophys. J. Lett.}, 918(2):L28, sep 2021.

\bibitem{FlanaganHinderer2008}
E.E. Flanagan and T.~Hinderer.
\newblock Constraining neutron-star tidal love numbers with gravitational-wave
  detectors.
\newblock {\em Phys. Rev. D}, 77(2):021502, 2008.

\bibitem{Hinderer2008}
T.~Hinderer.
\newblock Tidal love numbers of neutron stars.
\newblock {\em Astrophys. J.}, 677(2):1216, 2008.

\bibitem{BinningtonPoisson2009}
T.~Binnington and E.~Poisson.
\newblock Relativistic theory of tidal love numbers.
\newblock {\em Phys. Rev. D}, 80(8):084018, 2009.

\bibitem{DamourNagar2009}
T.~Damour and A.~Nagar.
\newblock Relativistic tidal properties of neutron stars.
\newblock {\em Phys. Rev. D}, 80(8):084035, 2009.

\bibitem{RadiceEtal2018}
D.~Radice, A.~Perego, F.~Zappa, and S.~Bernuzzi.
\newblock {GW}170817: Joint constraint on the neutron star equation of state
  from multimessenger observations.
\newblock {\em Astrophys. J.}, 852(2):L29, jan 2018.

\bibitem{HernandezNunezVasquez2018}
H.~Hern{\'a}ndez, L.~A. N{\'u}{\~n}ez, and A.~V{\'a}squez-Ram{\'\i}rez.
\newblock Convection and cracking stability of spheres in general relativity.
\newblock {\em Eur. Phys. J. C}, 78(11):883, 2018.

\bibitem{MisnerSharp1964}
C.~W. {Misner} and D.~H. {Sharp}.
\newblock {Relativistic Equations for Adiabatic, Spherically Symmetric
  Gravitational Collapse}.
\newblock {\em Phys, Rev.}, 136:571--576, October 1964.

\bibitem{Herrera1992}
L.~Herrera.
\newblock Cracking of self-gravitating compact objects.
\newblock {\em Phys. Lett. A}, 165(3):206--210, 1992.

\bibitem{DiPriscoHerreraVarela1997}
A.~Di~Prisco, L.~Herrera, and V.~Varela.
\newblock Cracking of homogeneous self-gravitating compact objects induced by
  fluctuations of local anisotropy.
\newblock {\em Gen. Relativ. Gravitation}, 29(10):1239--1256, 1997.

\bibitem{AbreuHernandezNunez2007b}
H.~Abreu, H.~Hern{\'a}ndez, and L.~A. N{\'u}{\~n}ez.
\newblock Sound speeds, cracking and stability of self-gravitating anisotropic
  compact objects.
\newblock {\em Class. Quantum Grav.}, 24(18):4631--4646, 2007.

\bibitem{GonzalezNavarroNunez2015}
G.~A. Gonz{\'a}lez, A.~Navarro, and L.~A. N\'u{\~n}ez.
\newblock Cracking of anisotropic spheres in general relativity revisited.
\newblock {\em J. Phys. Conf. Ser.}, 600(1):012014, 2015.

\bibitem{GonzalezNavarroNunez2017}
G.~A. {Gonz{\'a}lez}, A.~{Navarro}, and L.~A. {N{\'u}{\~n}ez}.
\newblock {Cracking isotropic and anisotropic relativistic spheres}.
\newblock {\em Can. J. Phys.}, 95(11):1089--1095, 2017.

\bibitem{Florides1974}
P.~S. Florides.
\newblock A new interior schwarzschild solution.
\newblock {\em Proceeding of the Royal Society of London}, A337:529 -- 535,
  1974.

\bibitem{DelgatyLake1998}
M.~S.~R. Delgaty and K.~Lake.
\newblock {Physical acceptability of isolated, static, spherically symmetric,
  perfect fluid solutions of Einstein's equations}.
\newblock {\em Comput. Phys. Commun.}, 115:395, 1998.

\bibitem{Buchdahl1959}
H.~A. Buchdahl.
\newblock General relativistic fluid spheres.
\newblock {\em Phys. Rev.}, 116(4):1027--1034, 1959.

\bibitem{Ivanov2002B}
B.~V. {Ivanov}.
\newblock {Maximum bounds on the surface redshift of anisotropic stars}.
\newblock {\em Phys. Rev. D}, 65(10):104011, 2002.

\bibitem{KolassisSantosTsoubelis1998}
C.~A. Kolassis, N.~O. Santos, and D.~Tsoubelis.
\newblock Energy conditions for an imperfect fluid.
\newblock {\em Class. Quantum Grav.}, 5(10):1329--1338, 1988.

\bibitem{PimentelLoraGonzalez2017}
O.~M. {Pimentel}, F.~D. {Lora-Clavijo}, and G.~A. {Gonz{\'a}lez}.
\newblock {Ideal magnetohydrodynamics with radiative terms: energy conditions}.
\newblock {\em Class. Quantum Grav.}, 34(7):075008, 2017.

\bibitem{HeintzmannHillebrandt1975}
H.~Heintzmann and W.~Hillebrandt.
\newblock {Neutron stars with an anisotropic equation of state: Mass, redshift
  and stability}.
\newblock {\em Astron. Astrophys.}, 38:51--55, 1975.

\bibitem{ChanHerreraSantos1993}
R.~Chan, L.~Herrera, and N.~O. Santos.
\newblock Dynamical instability for radiating anisotropic collapse.
\newblock {\em Mon. Not. R. Astron. Soc.}, 265(3):533--544, 1993.

\bibitem{ChanHerreraSantos1994}
R.~Chan, L.~Herrera, and N.O. Santos.
\newblock Dynamical instability for shearing viscous collapse.
\newblock {\em Mon. Not. R. Astron. Soc.}, 267(3):637--646, 1994.

\bibitem{HarrisonThorneWakano1965}
B.~K. {Harrison} et~al.
\newblock {\em Gravitation theory and gravitational collapse}.
\newblock University of Chicago Press, Chicago, 1965.

\bibitem{ZeldovichNovikov1971}
Y.~B. {Zeldovich} and I.~D. {Novikov}.
\newblock {\em {Relativistic astrophysics. Vol.1: Stars and relativity}}.
\newblock University of Chicago Press, Chicago, 1971.

\bibitem{DevGleiser2003}
K.~Dev and M.~Gleiser.
\newblock Anisotropic stars {II}: Stability.
\newblock {\em Gen. Relativ. Gravitation}, 35(8):1435--1457, 2003.

\bibitem{GleiserDev2004}
M.~Gleiser and K.~Dev.
\newblock Anistropic stars: Exact solutions and stability.
\newblock {\em Int. J. Mod. Phys. D}, 13(07):1389--1397,
  2004.

\bibitem{Bondi1964B}
H.~Bondi.
\newblock {Massive Spheres in General Relativity}.
\newblock {\em Proc. Math. Phys. Eng. Sci. P ROY SOC A-MATH PHY}, 282(1390):303--317, 1964.

\bibitem{Thorne1966}
K.~S. Thorne.
\newblock {Validity in General Relativity of the Schwarzschild Criterion for
  Convection}.
\newblock {\em Astrophys. J.}, 144:201--205, April 1966.

\bibitem{HerreraBarreto2013}
L.~Herrera and W.~Barreto.
\newblock General relativistic polytropes for anisotropic matter: The general
  formalism and applications.
\newblock {\em Phys. Rev. D}, 88(8):084022, 2013.

\bibitem{HerreraBarreto2013B}
L.~Herrera and W.~Barreto.
\newblock Newtonian polytropes for anisotropic matter: General framework and
  applications.
\newblock {\em Phys. Rev. D}, 87(8):087303, 2013.

\bibitem{HorvatIIijicMarunovic2011B}
D.~{Horvat}, S.~{Iliji{c}}, and A.~{Marunovi{c}}.
\newblock {Radial pulsations and stability of anisotropic stars with a
  quasi-local equation of state}.
\newblock {\em Class. Quantum Grav.}, 28(2):025009, January 2011.

\bibitem{Bludman1973}
S.~A. {Bludman}.
\newblock {Stability of General-Relativistic Polytropes}.
\newblock {\em Astrophys. J.}, 183:637--648, 1973.

\bibitem{Horedt2004}
G.~P. Horedt.
\newblock {\em Polytropes: applications in astrophysics and related fields},
  volume 306.
\newblock Springer Science \& Business Media, 2004.

\bibitem{GokhrooMehra1994}
M.~K. Gokhroo and A.~L. Mehra.
\newblock Anisotropic spheres with variable energy density in general
  relativity.
\newblock {\em Gen. Rel. Grav.}, 26(1):75 -- 84, 1994.

\bibitem{RileyEtal2019}
T.~E. Riley, A.~L. Watts, S.~Bogdanov, P.~S. Ray, R.~M. Ludlam, S.~Guillot,
  Z.~Arzoumanian, C.~L. Baker, A.~V. Bilous, D.~Chakrabarty, K.~C. Gendreau,
  A.~K. Harding, W.~C.~G. Ho, J.~M. Lattimer, S.~M. Morsink, and T.~E.
  Strohmayer.
\newblock A nicer view of psr j0030+0451: Millisecond pulsar parameter
  estimation.
\newblock {\em Astrophys. J.}, 887(1):L21, dec 2019.

\bibitem{ArrollochavezEtal2020}
G. Arroyo-Ch\'avez, A.~Cruz-Osorio, F.D. Lora-Clavijo, C. Campuzano-Vargas, and
  L.A. Garc\'ia-Mora.
\newblock Neutron and quark stars: constraining the parameters for simple eos
  using the gw170817.
\newblock {\em Astrophys. Space Sci.}, 365(2), 02 2020.

\bibitem{PoissonWill2014}
E.~Poisson and C.M. Will.
\newblock {\em Gravity: Newtonian, Post-Newtonian, Relativistic}.
\newblock Cambridge University Press, 2014.

\bibitem{YagiYunez2015}
K.~Yagi and N.~Yunes.
\newblock I-love-q anisotropically: Universal relations for compact stars with
  scalar pressure anisotropy.
\newblock {\em Phys. Rev. D}, 91(12):123008, 2015.

\bibitem{RaposoetEtal2019}
G.~Raposo, P.~Pani, M.~Bezares, C.~Palenzuela, and V.~Cardoso.
\newblock Anisotropic stars as ultracompact objects in general relativity.
\newblock {\em Phys.Rev.D}, 99:104072, 2019.

\bibitem{DasEtal2021}
S.~{Das}, S.~{Ray}, M.~{Khlopov}, K.~K. {Nandi}, and B.~K. {Parida}.
\newblock {Anisotropic compact stars: Constraining model parameters to account
  for physical features of tidal Love numbers}.
\newblock {\em Ann Phys (N Y)}, 433:168597, October 2021.

\bibitem{ArbanilPanotopoulos2022}
J.D.V. Arba{\~n}il and G.~Panotopoulos.
\newblock Tidal deformability and radial oscillations of anisotropic polytropic
  spheres.
\newblock {\em Phys. Rev. D}, 105(2):024008, 2022.

\bibitem{ChirentiPosadaGuedes2020}
C.~Chirenti, C.~Posada, and V.~Guedes.
\newblock Where is love? tidal deformability in the black hole compactness
  limit.
\newblock {\em Class. Quantum Gravity}, 37(19):195017, sep 2020.

\bibitem{BurgayEtal2003}
M.~Burgay et~al.
\newblock An increased estimate of the merger rate of double neutron stars from
  observations of a highly relativistic system.
\newblock {\em Nature}, 426(6966):531--533, 2003.

\bibitem{LyneEtal2004}
A.G. Lyne, M.~Burgay, M.~Kramer, A.~Possenti, R.N. Manchester, F.~Camilo, M.A.
  McLaughlin, D.R. Lorimer, N.~D'Amico, B.C. Joshi, J.~Reynolds, and C.C.
  Freire.
\newblock A double-pulsar system: a rare laboratory for relativistic gravity
  and plasma physics.
\newblock {\em Science}, 303(5661):1153--1157, 2004.

\bibitem{NiceSayerTaylor1996}
D.~J. Nice, R.~W. Sayer, and J.~H. Taylor.
\newblock {PSR} {J}1518+4904: A mildly relativistic binary pulsar system.
\newblock {\em Astrophys. J. Lett.}, 466(2):L87--L90, 1996.

\bibitem{JanssenEtal2008}
G.~H. {Janssen}, B.~W. {Stappers}, M.~{Kramer}, D.~J. {Nice}, A.~{Jessner},
  I.~{Cognard}, and M.~B. {Purver}.
\newblock Multi-telescope timing of {PSR} {J}1518+4904.
\newblock {\em Astron. Astrophys.}, 490(2):753--761, 2008.

\bibitem{SuarezurangoNunezHernandez2021}
D.~Su{\'a}rez-Urango, L.A. N{\'u}{\~n}ez, and H.~Hern{\'a}ndez.
\newblock Relativistic anisotropic polytropic spheres: Physical acceptability.
\newblock {\em arXiv:2102.00496}, 2021.

\end{thebibliography}

\end{document}